\def\sample{492}
\def\sampleb{535}
\def\samplebsvdone{152}
\def\samplebsvdtwo{383}

\def\samplebwithtimeinf{258}

\def\ifb{\ensuremath{\text{fb}^{-1}}}
\def\iab{\ensuremath{\text{ab}^{-1}}}

\def\BF{\mathcal{B}}
\def\Lsyst{\mathcal{L}_{\rm syst}}

\def\etap{\ensuremath{\eta^{\prime}}}
\def\etac{\ensuremath{\eta_{c}}}
\def\etacp{\ensuremath{\eta_{c}(2S)}}
\def\chicz{\ensuremath{\chi_{c0}}}
\def\chict{\ensuremath{\chi_{c2}}}
\def\jpsi{\ensuremath{J/\psi}}
\def\Xt{\ensuremath{X(3872)}}
\def\piz{\ensuremath{\pi^{0}}}
\def\UfourS{\ensuremath{\Upsilon(4S)}}

\def\BtoKgg{\ensuremath{B^{\pm}\to{}K^{\pm}\gamma\gamma}}
\def\BtoKhtoKgg{\ensuremath{B^{\pm}\to{} K^{\pm} h \to{} K^{\pm}\gamma\gamma}}

\def\BtoKeta{\ensuremath{B^{\pm}\to{}K^{\pm}\eta}}
\def\BtoKetatoKgg{\ensuremath{B^{\pm} \to{} K^{\pm}\eta \to{} K^{\pm}\gamma\gamma}}

\def\BtoKetap{\ensuremath{B^{\pm}\to{}K^{\pm}\etap}}
\def\BtoKetaptoKgg{\ensuremath{B^{\pm}\to{}K^{\pm}\etap \to{} K^{\pm}\gamma\gamma }}

\def\BtoKetac{\ensuremath{B^{\pm}\to{}K^{\pm}\etac}}
\def\BtoKetactoKgg{\ensuremath{B^{\pm}\to{}K^{\pm}\etac \to{} K^{\pm}\gamma\gamma }}

\def\htogg{\ensuremath{h \to \gamma \gamma}}
\def\BtoKh{\ensuremath{B^{\pm}\to{} K^{\pm} h}}

\def\BtoKetacp{\ensuremath{B^{\pm}\to{}K^{\pm}\etacp}}
\def\BtoKetacptoKgg{\ensuremath{B^{\pm}\to{}K^{\pm}\etacp \to{} K^{\pm}\gamma\gamma }}
\def\BtoKchicz{\ensuremath{B^{\pm}\to{}K^{\pm}\chicz}}
\def\BtoKchicztoKgg{\ensuremath{B^{\pm}\to{}K^{\pm}\chicz \to{} K^{\pm}\gamma\gamma}}
\def\BtoKchict{\ensuremath{B^{\pm}\to{}K^{\pm}\chict}}
\def\BtoKchicttoKgg{\ensuremath{B^{\pm}\to{}K^{\pm}\chict \to{} K^{\pm}\gamma\gamma}}
\def\BtoKjpsi{\ensuremath{B^{\pm}\to{}K^{\pm}\jpsi}}
\def\BtoKjpsitoKgg{\ensuremath{B^{\pm}\to{}K^{\pm}\jpsi \to{} K^{\pm}\gamma\gamma}}
\def\BtoKXt{\ensuremath{B^{\pm}\to{}K^{\pm}\Xt}}
\def\BtoKXttoKgg{\ensuremath{B^{\pm}\to{}K^{\pm}\Xt \to{} K^{\pm}\gamma\gamma}}

\def\BtoKpiz{\ensuremath{B^{\pm}\to{}K^{\pm}\piz}}

\def\BtoKstgamma{\ensuremath{B^{\pm}\to{}K^{*\pm}\gamma \to K^{\pm} \gamma\gamma}}

\def\BtoKsteta{\ensuremath{B \to K^{*} \eta}}
\def\BtoKstetac{\ensuremath{B \to K^{*} \etac}}

\def\etactogg{\ensuremath{\etac \to \gamma \gamma}}

\def\chicztogg{\ensuremath{\chicz \to \gamma \gamma}}
\def\jpsitogg{\ensuremath{\jpsi \to \gamma \gamma}}
\def\etacptogg{\ensuremath{\etacp \to \gamma \gamma}}

\def\mevct{\ensuremath{\text{MeV}/c^{2}}}
\def\gevct{\ensuremath{\text{GeV}/c^{2}}}

\def\mev{\text{MeV}}
\def\gev{\text{GeV}}

\def\mbc{\ensuremath{M_{\mathrm{bc}}}}
\def\deltae{\ensuremath{{\Delta}E}}

\newcommand{\bbar}{\ensuremath{B\overline{B}}}

\def\totalyieldBtoKeta{\ensuremath{76 {^{+14}_{-13}} }}
\def\brBtoKeta{\ensuremath{0.87 { ^{+0.16}_{-0.15}  } { ^{+0.10}_{-0.07}}}}
\def\brBtoKetass{\ensuremath{0.87 { ^{+0.16}_{-0.15}(\rm stat) } { ^{+0.10}_{-0.07} (\rm syst) }}}
\def\systsignifbrBtoKeta{\ensuremath{7.3}}

\def\totalyieldBtoKetap{\ensuremath{114 \pm 13 }}
\def\brBtoKetap{\ensuremath{1.40  { ^{+0.16}_{-0.15} } { ^{+0.15}_{-0.12} } }}
\def\brBtoKetapss{\ensuremath{1.40  { ^{+0.16}_{-0.15}(\rm stat) } { ^{+0.15}_{-0.12} (\rm syst) } }}
\def\systsignifbrBtoKetap{\ensuremath{13.8}}

\def\bretaptoggss{\ensuremath{2.01 {^{+0.23}_{-0.22}} {^{+0.23}_{-0.19}}}}

\def\totalyieldBtoKetac{\ensuremath{13.3 {^{+4.8}_{-4.1}} }}
\def\brBtoKetac{\ensuremath{0.22 { ^{+0.09}_{-0.07} } { ^{+0.04}_{-0.02} }}}
\def\brBtoKetacss{\ensuremath{0.22 { ^{+0.09}_{-0.07}(\rm stat) } { ^{+0.04}_{-0.02} (\rm syst) }}}
\def\systsignifbrBtoKetac{\ensuremath{4.1}}

\def\bretactogg{2.4 {^{+0.9}_{-0.8} (\rm stat)} {^{+0.7}_{-0.4}(\rm syst)} }
\def\bretactoggss{2.4 {^{+0.9}_{-0.8}} {^{+0.7}_{-0.4}} }

\def\totalyieldBtoKetacp{\ensuremath{4.0 {^{+3.9}_{-3.0}} }}

\def\systlimitbrBtoKetacp{0.18}

\def\systlimitbretacptogamgam{\ensuremath{8.2}}

\def\totalyieldBtoKchicz{\ensuremath{0.7 { ^{+2.5}_{-1.7}}}}

\def\systlimitbrBtoKchicz{0.11}

\def\systlimitbrchicztogamgam{9.5}

\def\totalyieldBtoKchict{\ensuremath{-0.3 {^{+2.6}_{-1.9} }  }}

\def\systlimitbrBtoKchict{0.09}

\def\totalyieldBtoKjpsi{\ensuremath{3.4 {^{+2.8}_{-2.0}} }}

\def\systlimitbrBtoKjpsi{0.16}

\def\systlimitbrjpsitogamgam{1.6}

\def\totalyieldBtoKXt{\ensuremath{-0.9 {^{+2.2}_{-1.4}} }}

\def\systlimitbrBtoKXt{0.24}

\def\systlimitbrXttogamgam{1.1}

\def\kbrBtoKeta{\ensuremath{ (2.6 \pm 0.6) \times 10^{-6} }}
\def\kbretatogg{\ensuremath{ (39.39 \pm 0.24) \% }}
\def\kbrBtoKetatoKgg{\ensuremath{ (1.02 \pm 0.24) \times 10^{-6} }}

\def\kbrBtoKetap{\ensuremath{ (69.7 \pm 2.8) \times 10^{-6} }}
\def\kbretaptogg{\ensuremath{ (2.12 \pm 0.14) \%  }}
\def\kbrBtoKetaptoKgg{\ensuremath{ (1.48 \pm 0.11) \times 10^{-6} }}

\def\kbrBtoKetac{\ensuremath{ (9.1 \pm 1.3) \times 10^{-4} }}
\def\kbretactogg{\ensuremath{ (2.7 \pm 0.9) \times 10^{-4}  }}
\def\kbrBtoKetactoKgg{\ensuremath{ (0.25 \pm 0.09) \times 10^{-6}  }}

\def\kbrBtoKetacp{\ensuremath{ (3.4 \pm 1.8) \times 10^{-4} }}
\def\kbretacptogg{seen}
\def\kbrBtoKetacptoKgg{\ensuremath{  }}

\def\kbrBtoKchicz{\ensuremath{ (1.40 {^{+0.23}_{-0.19}}) \times 10^{-4}}}
\def\kbrchicztogg{\ensuremath{ (2.76 \pm 0.33) \times 10^{-4} }}
\def\kbrBtoKchicztoKgg{\ensuremath{ (0.39 \pm 0.08) \times 10^{-7} }}

\def\kbrBtoKchict{\ensuremath{< 2.9 \times 10^{-5}}}
\def\kbrchicttogg{\ensuremath{ (2.58 \pm 0.19) \times 10^{-4} }}
\def\kbrBtoKchicttoKgg{\ensuremath{ < 7.5 \times 10^{-9} }}

\def\kbrBtoKjpsi{\ensuremath{ (10.07 \pm 0.35) \times 10^{-4} }}
\def\kbrjpsitogg{\ensuremath{ < 9.3 \times 10^{-5} }}
\def\kbrBtoKjpsitoKgg{\ensuremath{ < 9.4 \times 10^{-8}   }}

\def\kbrBtoKXt{seen}
\def\kbrXttogg{\ensuremath{        }}
\def\kbrBtoKXttoKgg{\ensuremath{   }}

\documentclass{elsart}
\journal{Physics Letters B}
\usepackage{graphicx}
\usepackage{amssymb}
\usepackage{amsmath}
\usepackage{hyperref}

\usepackage[dvips]{color} 
\usepackage[figure,table]{hypcap}

\begin{document}

\begin{frontmatter}

\begin{flushleft}
\hspace*{9cm}BELLE Preprint 2007-22 \\
\hspace*{9cm}KEK Preprint 2007-11 \\
\end{flushleft}

\title{ Search for Resonant $B^{\pm}\to K^{\pm} h \to K^{\pm} \gamma \gamma$ Decays at Belle }

\collab{Belle Collaboration}
\author[Lausanne]{J.~Wicht}, 
  \author[KEK]{I.~Adachi}, 
  \author[Tokyo]{H.~Aihara}, 
  \author[BINP]{D.~Anipko}, 
  \author[BINP]{V.~Aulchenko}, 
  \author[Lausanne,ITEP]{T.~Aushev}, 
  \author[Sydney]{A.~M.~Bakich}, 
  \author[Melbourne]{E.~Barberio}, 
  \author[BINP]{I.~Bedny}, 
  \author[JSI]{U.~Bitenc}, 
% ...

\address[BINP]{Budker Institute of Nuclear Physics, Novosibirsk, Russia}
% ...
\address[KEK]{High Energy Accelerator Research Organization (KEK), Tsukuba, Japan}
% ...
\address[ITEP]{Institute for Theoretical and Experimental Physics, Moscow, Russia}
\address[JSI]{J. Stefan Institute, Ljubljana, Slovenia}
% ...
\address[Lausanne]{\'Ecole Polytechnique F\'ed\'erale de Lausanne (EPFL), Lausanne, Switzerland}
% ...
\address[Melbourne]{University of Melbourne, Victoria, Australia}
% ...
\address[Sydney]{University of Sydney, Sydney, NSW, Australia}
% ...
\address[Tokyo]{Department of Physics, University of Tokyo, Tokyo, Japan}
% ...

\begin{abstract}
We report measurements and searches for resonant \BtoKhtoKgg\ decays where $h$ is a $\eta,\eta^{\prime},\eta_{c},\eta_{c}(2S),\chi_{c0},\chi_{c2},J/\psi$ 
meson or the $X(3872)$ particle. The results are based on a data sample containing \sampleb\ million \bbar\ pairs collected with the Belle 
detector at the KEKB $e^+e^-$ asymmetric-energy collider operating at the \UfourS\ resonance. Signals are observed in the modes with $\eta$ and \etap, and we obtain evidence for a signal in the mode with \etac. We measure $\BF(\BtoKetatoKgg) = (\brBtoKetass) \times 10^{-6}$, $\BF(\BtoKetaptoKgg)$ $=$ $(\brBtoKetapss) \times 10^{-6}$ and $\BF(\BtoKetactoKgg)$ = \newline $(\brBtoKetacss) \times 10^{-6}$. We set upper limits on the branching fractions of the other modes.
\end{abstract}


\end{frontmatter}

\section{Introduction}

We report searches for resonant 
$B^{\pm} \rightarrow K^{\pm} h \rightarrow K^{\pm} \gamma \gamma$ decays, where $h$ can be one of the following mesons: $\eta$, \etap, \etac, \etacp, \chicz, \chict, \jpsi\ or the $X(3872)$~\cite{x3872-belle,x3872-cdf,x3872-d0,x3872-babar} particle. 

The nature and quantum numbers of the \Xt\ particle are still being debated; based on analyses of the dipion mass spectrum~\cite{x3872-1ppor2pp-belle,x3872-1pp-belle} and angular distributions~\cite{x3872-1ppor2pp-belle,x3872-1pp2mp-cdf} for $X(3872) \to \pi^{+} \pi^{-} J/\psi$, $J^{PC} = 1^{++}$ and $2^{-+}$ are allowed.
The $1^{++}$ assignment is also supported by signals observed in $B \to (D^{0} \overline{D}{}^{0} \pi^{0}) K$~\cite{x3872-1pp-belle} and in $B \to (D^{*0} \overline{D}{}^{0}) K$~\cite{x3872-dstd-babar} under the assumption that they are indeed due to the \Xt\ particle. The observation of $X(3872) \to J/\psi \rho$~\cite{x3872-ceq1-cdf} and $X(3872) \to J/\psi \gamma$~\cite{x3872-ceq1-belle,x3872-ceq1-babar} indicates that $C=+1$. Evidence of a signal in the \BtoKXttoKgg\ channel would rule out $J = 1$ since the decay of a spin $1$ particle (here the \Xt) into two photons is forbidden by gauge invariance and Bose-Einstein statistics~\cite{jpsiforbidden}.

Many of the \BtoKh\ and \htogg\ branching fractions involved in these decay chains have been already measured, as shown in Table~\ref{table:whatsknown}. The \BtoKeta\ and \BtoKetap\ modes are well established~\cite{PDG2006} and can be used as calibrations in the search for other \BtoKhtoKgg\ channels that have lower or unknown branching fractions. The \BtoKjpsi\ channel can also serve as a control mode, since the \jpsi\ is a spin $1$ particle and cannot decay into two photons.

The interference of \BtoKetactoKgg\ or \BtoKchicztoKgg\ with the radiative decay chain \BtoKstgamma\ can be used to measure the photon polarization in the $b \to s\gamma$ quark transition~\cite{schietinger}. Such measurement would provide a test of the Standard Model, which predicts the photon to be predominantly left-handed in $b \to s\gamma$ decays and right-handed in $\bar{b} \to{} \bar{s} \gamma$ decays. The observation of the \BtoKetactoKgg\ or \BtoKchicztoKgg\ decay chain is the first step in this search for new physics, which could be achieved with about 10 \iab\ of data (thus requiring a Super $B$ factory~\cite{superbelle,superb}). The non-resonant decay \BtoKgg\ is very rare, with a branching fraction estimated to be of order $10^{-9}$~\cite{b2kgg} with a large background over the whole $m_{\gamma\gamma}$ phase-space from the resonant \BtoKstgamma\ channel~\cite{schietinger}.

In this study, we use a data sample of \sample\ \ifb\ containing $\sampleb \times 10^6$ \bbar\ pairs that were collected with the Belle detector at the KEKB asymmetric-energy $e^+e^-$ (3.5 on 8~\gev) collider~\cite{KEKB} operating at the \UfourS\ resonance.

The Belle detector is a large-solid-angle magnetic
spectrometer that
consists of a silicon vertex detector (SVD),
a 50-layer central drift chamber (CDC), an array of
aerogel threshold Cherenkov counters (ACC), 
a barrel-like arrangement of time-of-flight
scintillation counters (TOF), and an electromagnetic calorimeter
comprised of CsI(Tl) crystals (ECL) located inside 
a superconducting solenoid coil that provides a 1.5~T
magnetic field.  An iron flux-return located outside
the coil is instrumented to detect $K_L^0$ mesons and to identify
muons (KLM).  The detector
is described in detail elsewhere~\cite{Belle}.
Two inner detector configurations were used. A 2.0 cm beampipe
and a 3-layer silicon vertex detector was used for the first sample
of $\samplebsvdone \times 10^6 \bbar$ pairs (SVD1), while a 1.5 cm beampipe, a 4-layer
silicon detector and a small-cell inner drift chamber were used to record  
the remaining $\samplebsvdtwo \times 10^6 \bbar$ pairs (SVD2~\cite{svd2}).

\begin{table}
\centering
\caption{Current status of the measured branching fractions or 90\% confidence level upper limits for \BtoKh\ and \htogg\ (all values are taken from Ref.~\cite{PDG2006}, unless otherwise indicated). The values in the last column are the expectations computed as the products $\BF(\BtoKh)\times \BF(\htogg)$. The decay chain \BtoKhtoKgg\ has only been observed for $h=\eta$.}

\begin{tabular}{l  c  c  c }\hline
$h$ & $\BF(B^{\pm} \! \to \! K^{\pm} h)$                      &  $\BF(h \! \to \! \gamma \gamma)$ & \hspace{-3mm}  $\BF(B^{\pm} \! \to K^{\pm} h \! \to \! K^{\pm} \gamma \gamma)$       \\ \hline 
$\eta$    & \kbrBtoKeta    & \kbretatogg                 & \kbrBtoKetatoKgg      \\ 
\etap     & \kbrBtoKetap   & \kbretaptogg                & \kbrBtoKetaptoKgg      \\
\etac     & \kbrBtoKetac   & \kbretactogg                & \kbrBtoKetactoKgg      \\ 
\etacp    & \kbrBtoKetacp  & \kbretacptogg               & \kbrBtoKetacptoKgg     \\ 
\chicz    & \kbrBtoKchicz  & \kbrchicztogg               & \kbrBtoKchicztoKgg      \\
\chict    & \kbrBtoKchict  & \kbrchicttogg               & \kbrBtoKchicttoKgg      \\
\jpsi     & \kbrBtoKjpsi   & \kbrjpsitogg~\cite{jpsiul}  & \kbrBtoKjpsitoKgg     \\ 
\Xt & \kbrBtoKXt~\cite{x3872-belle} & \kbrXttogg & \kbrBtoKXttoKgg     \\ \hline
\end{tabular}
\label{table:whatsknown}
\end{table}

\section{Event selection and background rejection}

Kaon candidates are selected from charged tracks with the requirement $\mathcal{L} = \mathcal{L}_K/(\mathcal{L}_K+\mathcal{L}_\pi)>0.6$, where $\mathcal{L}_{K}$ ($\mathcal{L}_{\pi}$) is the likelihood for a track to be a kaon (pion) based on the 
response of the ACC and on measurements from the CDC and TOF. 
The kaon identification efficiency is between $84\%$ and $90\%$ depending on the $K \gamma \gamma$ signal mode with 7\%--11\% of pions misidentified as kaons.
Photon pairs are selected 
by requiring their energies in the laboratory frame to be greater than 100~\mev\ and their 
energy asymmetry 
$A_{\gamma\gamma} = |\frac{E_{\gamma 1}-E_{\gamma 2}}{E_{\gamma 1}+E_{\gamma 2}}|$  
to be less than $0.9$. We reject photons from $\pi^0$ decays by removing photon pairs with an invariant mass between 117.8~\mevct\ and 150.2~\mevct\ (2.5 standard deviations around the $\pi^{0}$ mass). We require a shower shape consistent with that of a photon: for each cluster, the ratio of the energy deposited in the array of the central $3 \times 3$ calorimeter cells to that of $5 \times 5$ cells is computed. The cluster associated with the most energetic photon of the candidate pair is required to have a ratio greater than 0.95 while the cluster from the other photon must have a ratio greater than 0.90 for the \BtoKeta\ and \BtoKetap\ channels and 0.95 for the other channels.

Pairs of photons are retained and associated to the corresponding meson when
their invariant mass ($m_{\gamma\gamma}$) is inside one of the \emph{wide} mass windows defined in 
Table~\ref{table:mggsignalsel}. A mass-constrained fit of the photon momenta is performed to match 
the nominal~\cite{PDG2006} masses with the constraint that the photons originate from the interaction 
point. 

\begin{table}
\centering
\caption{Nominal mass [\gevct] of the reconstructed particles and definition of invariant mass windows [\gevct] for photon pairs.}
\begin{tabular*}{\textwidth}{@{\extracolsep{\fill}}l  c  c  c}\hline
  Particle & Mass & Wide $m_{\gamma\gamma}$ window & Tight $m_{\gamma\gamma}$ window  \\ \hline
  $\eta$   & 0.548        & 0.4--0.7                       &  0.50--0.57   \\
  \etap    & 0.958        & 0.8--1.1                       &  0.90--0.98   \\
  \etac    & 2.980        & 2.5--3.2                       &  2.82--3.05   \\
  \etacp   & 3.637        & 3.2--3.8                       &  3.44--3.70   \\
  \chicz   & 3.415        & 3.0--3.5                       &  3.25--3.50   \\
  \chict   & 3.556        & 3.0--3.8                       &  3.40--3.62   \\
  \jpsi    & 3.097        & 2.5--3.2                       &  2.92--3.15   \\
  \Xt      & 3.872        & 3.0--4.1                       &  3.72--3.95   \\ \hline
\end{tabular*}
\label{table:mggsignalsel}
\end{table}

Charged $B$ meson candidates are reconstructed starting from a kaon and a pair of photons, 
and they are selected by means of the beam-energy constrained mass, defined as 
$\mbc = \sqrt{E_\mathrm{beam}^{*2} - p_{B}^{*2}}$ and the energy difference 
$\deltae = E_{B}^{*} - E_\mathrm{beam}^{*}$.
In these definitions, $E_\mathrm{beam}^{*}$ is the beam energy and 
$p_{B}^{*}$ and $E_{B}^{*}$ are the momentum and the energy of the $B$ meson, all variables 
being evaluated in the center-of-mass (CM) frame. 
We select $B$-meson candidates with $\mbc > 5.2\;\gevct$ 
and $-0.3\;\gev < \deltae < 0.3\;\gev$. 
If more than one $B$ candidate is reconstructed in an event, 
the best candidate is chosen by selecting 
the photon pair with the smallest $\chi^{2}$ of the mass fit, and if multiple kaons can 
be associated with this photon pair, the kaon with the highest $\mathcal{L}$ is chosen.

The main background in all modes is due to continuum events, i.e.\ events coming from light-quark 
pair production ($u\bar{u}$, $d\bar{d}$, $s\bar{s}$ and $c\bar{c}$). The rejection of the continuum is studied and optimized using a Monte Carlo (MC) sample having about 1.5 times the size of the data sample.
Four variables are used to separate signal from continuum background: a Fisher discriminant 
based on modified Fox-Wolfram moments~\cite{SFW}, the $B$ production angle with respect to 
the beam in the CM frame, $\cos\theta^{*}$, the flight length difference along the beam 
axis between the two $B$ mesons, and the flavor tagging information~\cite{TaggingNIM}. 
The Fisher discriminant, the $B$ production angle and the flight length difference are
combined into a likelihood ratio $LR = L_{s}/(L_{s} + L_{udsc})$, where $L_{s}$ and $L_{udsc}$ are the product of probability density functions (PDFs) of these variables for signal and continuum events. We use different $LR$ cuts depending on the flavor 
tagging information. The continuum rejection is achieved by simultaneously optimizing the $LR$ and $m_{\gamma\gamma}$ cuts (\emph{tight} $m_{\gamma\gamma}$ window in Table~\ref{table:mggsignalsel}) in order to maximize the figure of merit in the signal windows ($\mbc > 5.27~\gevct$ and $\deltae$ as described in Table~\ref{table:signalwindows}). The figure of merit is defined as $\mathcal{S} = N_{\rm s}/\sqrt{N_{\rm s}+N_{udsc}}$ for the \BtoKeta\ and \BtoKetap\ modes and $\epsilon/\sqrt{N_{udsc}}$ for all the other modes, where $N_{\rm s}$ and $N_{udsc}$ are the expected number of 
signal and continuum events and $\epsilon$ is the signal efficiency. The expected numbers of events are computed for an integrated
luminosity of \sample~\ifb\ and assuming the measured branching fractions~\cite{PDG2006}.

\begin{table}
\centering
\caption{Definition of the \deltae\ signal windows [\gev]. The \mbc\ signal windows are defined as $\mbc > 5.27~\gevct$ for all modes.}
\begin{tabular}{l  c  |l  c  }\hline
  Particle &  \deltae\ window & Particle &  \deltae\ window \\ \hline
  $\eta$    &  $-0.15 < \deltae < 0.10$ & \chicz   &  $-0.10 < \deltae < 0.10$ \\
  \etap     &  $-0.15 < \deltae < 0.10$ & \chict   &  $-0.06 < \deltae < 0.06$ \\
  \etac     &  $-0.10 < \deltae < 0.10$ & \jpsi    &  $-0.09 < \deltae < 0.09$ \\
  \etacp    &  $-0.08 < \deltae < 0.06$ & \Xt      &  $-0.09 < \deltae < 0.09$ \\ \hline
\end{tabular}
\label{table:signalwindows}
\end{table}

Exclusive backgrounds from charmless $B$ decays are studied using large MC samples having about 36 times the size of the data sample. 
In the \BtoKeta\ channel, 56\% of this type of background is from $B \to K^{*} \eta$ 
with the rest being composed of several small contributions, the largest ones being due to
$B \to X_s \gamma$ and $B^{\pm} \to \eta \pi^{\pm}$. In the \BtoKetap\ channel, the
dominant source (about 2/3) is 
$B \to X_s \gamma$, about half of which is from $B \to K^{*}(K \pi) \gamma$. 
For the other modes, about 95\% of the charmless $B$ decay contributions is due to $B \to X_s \gamma$. The final state with $K^{\pm} \pi^{0} \gamma$ is a significant background for modes with charmonia and with the $X(3872)$ resonance. 
It is suppressed by the requirement $m_{K\gamma_{2}} > 1.5\;\gevct$, where $m_{K\gamma_{2}}$ is the invariant mass of the system formed by the kaon and the lowest energy photon (in the laboratory frame) forming the $K\gamma\gamma$ candidate. For the \BtoKetac\ channel, the
$B \to K^{*}(892) \eta_c(\gamma\gamma)$ background is the most relevant contribution.

Another source of background is produced by the overlap of a hadronic event with a previous QED interaction (mainly Bhabha scattering) that has left energy deposits in the calorimeter. This off-time background is removed by using the timing information of the calorimeter clusters corresponding to each photon candidate. This timing information is only available for 
the most recent data, containing $\samplebwithtimeinf \times 10^{6}$ \bbar\ pairs. For the rest of the 
data, we include the background in the fit described in the following 
section, by modeling 
it according to the off-time background events rejected from 
the most recent data.

The tight $m_{\gamma\gamma}$ windows overlap for some of the $h$ decays, e.g.\ the mass window for \etac\ includes some candidates for \jpsi\ and vice versa. Dedicated studies have
shown that, for the dataset considered in this analysis, the only non-negligible cross-feed is due to \BtoKetac\ events
that are reconstructed in the \BtoKjpsi\ mode. This effect is included in the fit as described below.

\section{Fitting procedure and results}

We perform a two-dimensional unbinned extended maximum likelihood fit to \mbc\
and \deltae. The signals are described using PDFs modeled with the product of a Crystal Ball function~\cite{crystalball}
for \mbc\ and three Gaussian functions for \deltae, while the continuum background is modeled
with an ARGUS function~\cite{argus} for \mbc\ and a first order polynomial function for \deltae.
The effect of neglecting the correlation between \mbc\ and \deltae\ has been studied 
using MC signal samples embedded in toy continuum samples; the number of signal events 
returned by the fit is found to be 1-3\% smaller than the true number, depending on the 
$h$ mode. We take this bias into account by correcting the signal efficiencies and
adding a systematic uncertainty. Table~\ref{table:signaleff} lists the corrected efficiencies obtained for 
each mode in the two sub-samples with different inner detector configurations. The signal 
PDF parameters are determined on MC signal events. The \mbc\ resolution and the \deltae\ 
resolution and mean are then corrected using a control sample of \BtoKpiz\ events. The \bbar \ and off-time backgrounds are modeled with two-dimensional
KEYS~\cite{keyspdf} PDFs  
extracted from MC events and from the off-time data sample, respectively.
The normalizations of the \bbar \ and off-time backgrounds are fixed in the fit. 
For the  \BtoKjpsi\ mode, the $K^\pm \eta_c$ cross-feed is included with normalization fixed
to the value obtained in the corresponding signal fit.

\begin{table}
\centering
\caption{Signal efficiencies for the two configurations of the detector.}
\begin{tabular}{l c c | l c c }\hline
  Particle  & $\epsilon({\rm SVD1})$ [\%] & $\epsilon({\rm SVD2})$ [\%] &  Particle  & $\epsilon({\rm SVD1})$ [\%] & $\epsilon({\rm SVD2})$ [\%]\\ \hline
  $\eta$    & $15.8 \pm 0.1$         & $16.6 \pm 0.1$  &   \chicz    & $11.0 \pm 0.1$         & $11.6 \pm 0.1$  \\ 
  \etap      & $14.6 \pm 0.1$         & $15.7 \pm 0.1$ &    \chict    & $10.4 \pm 0.1$         & $11.3 \pm 0.1$  \\ 
  \etac      & $10.0 \pm 0.1$         & $10.9 \pm 0.1$ &   \jpsi      & $9.4 \pm 0.1$          & $9.7 \pm 0.1$  \\ 
  \etacp    & $10.9 \pm 0.1$         & $11.4 \pm 0.1$  &   \Xt         & $13.7 \pm 0.1$         & $15.0 \pm 0.1$  \\ \hline
\end{tabular}
\label{table:signaleff}
\end{table}

The fit is performed for \mbc\ greater than $5.2~\gevct$ and for \deltae\ between $-0.3~\gev$ and $0.3~\gev$. The likelihood is defined as: 
\begin{equation}
\mathcal{L} = e^{-\sum_{j}{N_j}} \times \prod_{i}(\sum_{j}{N_j P_j^i(\mbc^i,\deltae^i)})
\end{equation}
where $i$ runs over all events, $j$ runs over the possible event categories (signal, continuum background and other backgrounds), $N_j$ is the number of events in each category and $P_j$ is the corresponding PDF.

The data are divided into sub-samples based on the SVD configuration and the availability of the timing information needed for the rejection of off-time background.

The fit variables are the branching fraction ($\BF$) and the continuum background normalization and PDF parameters, except the ARGUS endpoint which is fixed to $E_{\rm beam}^{*} = 5.289~\gev$. 
The number of signal events is then defined as 
$S^{k}  = \BF \times \epsilon^{k} \times N_{\bbar}^{k}$ where $N_{\bbar}^{k}$ is the number of \bbar\ events and $\epsilon^{k}$ is the 
signal efficiency, both evaluated for sub-sample $k$.

The branching fraction obtained from the fit depends on the following parameters that can give rise to systematic uncertainties:
\begin{enumerate}
\item parameters related to particle reconstruction and identification and to signal selection, which affect the signal in a very similar way for all $h$ modes, as summarized in Table~\ref{table:systerrorsensitivity},
\item signal PDF parameters (0--5\% uncertainty),
\item normalization of the charmless $B$ and off-time backgrounds and of the $K^\pm \eta_c$ cross-feed for the \BtoKjpsi\ mode (1--10\%),
\item number of \bbar\ events (1.3\%).
\end{enumerate}

\begin{table}
\centering
\caption{Systematic uncertainties on the signal reconstruction efficiency.}
\begin{tabular}{l c }\hline
Source & Uncertainty [\%]  \\ \hline
Photon reconstruction efficiency & $2 \times 2.2$ \\ 
Tracking efficiency & $1 \times 1$ \\ 
Kaon identification efficiency & $2$ \\  
$m_{\gamma\gamma}$ cut efficiency & $3.6$ \\ 
$LR$ cut efficiency & $6.9$ \\
MC statistics      & $1.0$ \\
Fit bias           & $0.5$ \\ \hline 
Total               & $9.3$ \\ \hline
\end{tabular}
\label{table:systerrorsensitivity}
\end{table}

Systematic uncertainties related to the $m_{\gamma\gamma}$ and $LR$ requirements are evaluated by comparing efficiencies in data and MC using a \BtoKpiz\ control sample. Systematic uncertainties are included in the likelihood function by integration. The statistical likelihood is convolved with the probability distribution of the systematics parameters listed above, computed as the product of Gaussian terms, one for each parameter.
A MC integration is performed over the phase space of the systematics parameters, yielding a new likelihood function, $\Lsyst$, that includes all systematic uncertainties.
The fit results quoted below are all extracted from $\Lsyst$. The central value $\BF_0$ is the $\BF$ at which $\Lsyst$ has its maximum and the errors $\delta_{\rm tot}^{\pm}$ are defined by :
\begin{equation}
\frac{\int_{\BF_0+\delta_{\rm tot}^{-}}^{\BF_0+\delta_{\rm tot}^{+}} \Lsyst\,d\BF}{\int_{0}^{1} \Lsyst\,d\BF} = 0.68
\end{equation}
where the integration interval is chosen such that all points outside the interval have a lower likelihood than those inside. The positive (negative) systematic error is computed as $\pm \sqrt{{\delta_{\rm tot}^{\pm}}^2 - {\delta_{\rm stat}^{\pm}}^2 }$ where $\delta_{\rm stat}^{\pm}$ is the positive (negative) statistical error. The significance of the measurement of the branching fraction is defined as $\sqrt{2 (\ln{\Lsyst(\BF = \BF_0)-\ln{\Lsyst(\BF = 0)}})}$.
For modes in which no significant signal is found, 
the 90\% credible upper limit, $\BF_{\rm limit}$, is computed using a Bayesian approach
with a flat prior, according to:
\begin{equation}
\frac{\int_{0}^{\BF_{\rm limit}} \Lsyst\,d\BF}{\int_{0}^{1} \Lsyst\,d\BF} = 0.9 
\end{equation}

The fit results for all modes are summarized in Table~\ref{table:results}. We observe signals in the \BtoKeta\ and \BtoKetap\ modes and obtain evidence for a signal in the \BtoKetac\ channel, while we see no signal in the other modes. We report the first measurements of \BtoKetap\ and \BtoKetac\ channels in the $K^{\pm} \gamma \gamma$ final state. We measure $\BF(\BtoKetatoKgg) = (\brBtoKetass) \times 10^{-6}$ in agreement with Belle's measurement of this mode with the same
dataset~\cite{b2keta-belle}, $\BF(\BtoKetaptoKgg) = (\brBtoKetapss) \times 10^{-6}$ and $\BF(\BtoKetactoKgg) = (\brBtoKetacss) \times 10^{-6}$. All measured branching fractions agree with the values shown in the third column of Table~\ref{table:whatsknown}. Fit projections are shown in Figures~\ref{figure:data_eta} and \ref{figure:data_chicz}; in each plot the variable that is not shown is restricted to be in the signal window.


For the modes where no significant signal is observed, we extract the following 90\% probability upper limits:
 $\BF(\BtoKetacptoKgg) < \systlimitbrBtoKetacp \times 10^{-6}$, $\BF(\BtoKchicztoKgg) < \systlimitbrBtoKchicz \times 10^{-6}$, $\BF(\BtoKchicttoKgg) < \systlimitbrBtoKchict \times 10^{-6}$, $\BF(\BtoKjpsitoKgg) < \systlimitbrBtoKjpsi \times 10^{-6}$ and $\BF(\BtoKXttoKgg) < \systlimitbrBtoKXt \times 10^{-6}$. 
Whenever the branching fraction of \BtoKh\ has been measured elsewhere, we also perform the fit by constraining $\BF(\BtoKh)$ to the measured value~\cite{PDG2006}, thus extracting an upper limit on $\BF(\htogg)$. The uncertainty on $\BF(\BtoKh)$ is included as a source of systematic uncertainty. We obtain 
 $\BF(\chicztogg) < \systlimitbrchicztogamgam \times 10^{-4}$, $\BF(\etacptogg) < \systlimitbretacptogamgam \times 10^{-4}$ and
$\BF(\jpsitogg) < \systlimitbrjpsitogamgam \times 10^{-4}$ at 90\% probability.
Similarly, for the \BtoKetac\ mode, we determine $\BF(\etactogg) = (\bretactogg) \times 10^{-4}$.

The absolute branching fraction $\BF(\BtoKXt)$ has not yet been measured. However, there are measurements of the product of this quantity and the branching fractions of different decays of the $X(3872)$. Assuming that $X(3872)$ decays to $J/\psi \pi^+ \pi^-$, $J/\psi \pi^+ \pi^- \pi^0$ and $J/\psi \gamma$ saturate all possible decays of the $X(3872)$ and taking the values of the corresponding products from~\cite{PDG2006,x3872-ceq1-belle,x3872-ceq1-babar}, we derive a conservative upper limit $\BF(X(3872) \to \gamma \gamma) < \systlimitbrXttogamgam \%$ at 90\% probability. 

\begin{table}
\centering
\caption{Signal yields, branching fractions and significances ($\mathcal{S}$) results for \BtoKhtoKgg. The first uncertainty is statistical, the second one is systematic. Limits are calculated at 90\% probability.}
\begin{tabular*}{\textwidth}{@{\extracolsep{\fill}}lcccc}\hline 
Resonance   & Yield        & $\BF(\BtoKhtoKgg)$ ($10^{-6}$) & $\mathcal{S}$ & $\BF(\htogg)$          \\ \hline 
$\eta$ & \totalyieldBtoKeta       & \brBtoKeta        & \systsignifbrBtoKeta &  --  \\
\etap  & \totalyieldBtoKetap      & \brBtoKetap      & \systsignifbrBtoKetap &  $(\bretaptoggss) \%$ \\
\etac  & \totalyieldBtoKetac      & \brBtoKetac      & \systsignifbrBtoKetac &  $(\bretactoggss) \times 10^{-4}$  \\ 
\etacp & \phantom{-}\totalyieldBtoKetacp & $<$ \systlimitbrBtoKetacp & -- &  $ < \systlimitbretacptogamgam \times 10^{-4}$           \\ 
\chicz & \phantom{-}\totalyieldBtoKchicz & $<$ \systlimitbrBtoKchicz & -- &  $< \systlimitbrchicztogamgam \times 10^{-4}$          \\ 
\chict & \totalyieldBtoKchict & $<$ \systlimitbrBtoKchict            & --  &    --        \\ 
\jpsi  & \phantom{-}\totalyieldBtoKjpsi  & $<$ \systlimitbrBtoKjpsi  & -- &  $< \systlimitbrjpsitogamgam  \times 10^{-4}$        \\ 
\Xt    & \totalyieldBtoKXt    & $<$ \systlimitbrBtoKXt               & -- &  $< \systlimitbrXttogamgam \%$          \\ \hline
\end{tabular*}
\label{table:results}
\end{table}

\section{Conclusions}

A search for resonant \BtoKhtoKgg\ decays, where the resonance $h$ can be $\eta,\etap,\etac,\etacp,\chicz,\chict,\jpsi$ or $X(3872)$,  has been performed in a sample containing \sampleb\ million \bbar\ pairs. We have observed \BtoKeta\ and \BtoKetap\ with significances of \systsignifbrBtoKeta\ and \systsignifbrBtoKetap, respectively, and we have obtained evidence for \BtoKetac\ with a significance of \systsignifbrBtoKetac. No evidence of a signal is observed in any of the other modes and 90\% probability upper limits are set on the corresponding branching fractions. The measured branching fraction for \BtoKetatoKgg\ is in agreement with Belle's measurement of this mode with the same dataset~\cite{b2keta-belle}. We report the first observation of \BtoKetap\ and the first evidence of \BtoKetac\ in the $K^{\pm} \gamma \gamma$ final state.

We thank the KEKB group for the excellent operation of the
accelerator, the KEK cryogenics group for the efficient
operation of the solenoid, and the KEK computer group and
the National Institute of Informatics for valuable computing
and Super-SINET network support. We acknowledge support from
the Ministry of Education, Culture, Sports, Science, and
Technology of Japan and the Japan Society for the Promotion
of Science; the Australian Research Council and the
Australian Department of Education, Science and Training;
the National Science Foundation of China under
contract No.~10575109 and 10775142; the Department of
Science and Technology of India; 
the BK21 program of the Ministry of Education of Korea, 
the CHEP SRC program and Basic Research program 
(grant No.~R01-2005-000-10089-0) of the Korea Science and
Engineering Foundation, and the Pure Basic Research Group 
program of the Korea Research Foundation; 
the Polish State Committee for Scientific Research; 
the Ministry of Education and Science of the Russian
Federation and the Russian Federal Agency for Atomic Energy;
the Slovenian Research Agency;  the Swiss
National Science Foundation; the National Science Council
and the Ministry of Education of Taiwan; and the U.S.\
Department of Energy.

\clearpage

\begin{figure}
\centering
 \begin{tabular}{ll}
   \includegraphics[width=0.30\textwidth]{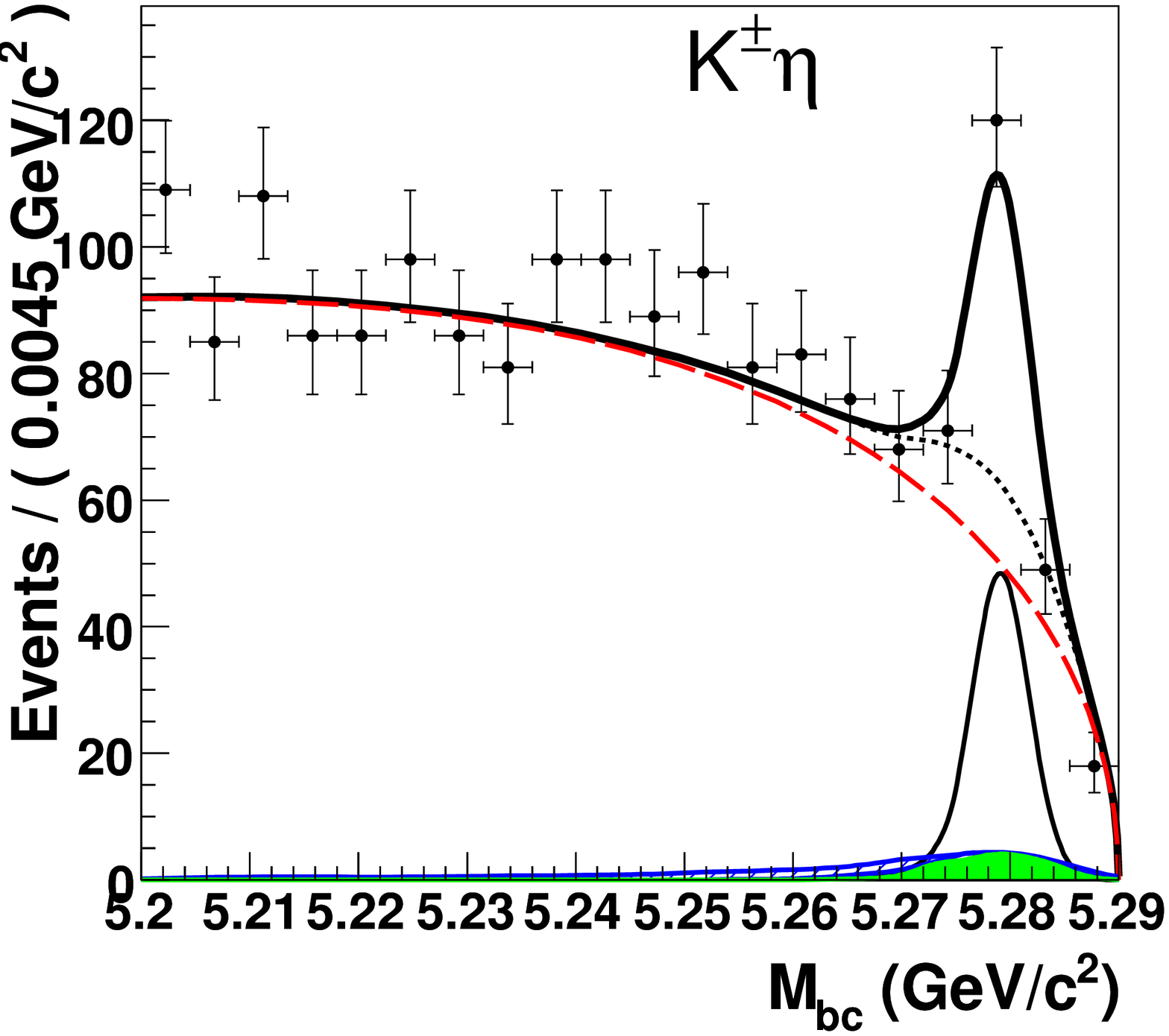} & 
   \includegraphics[width=0.30\textwidth]{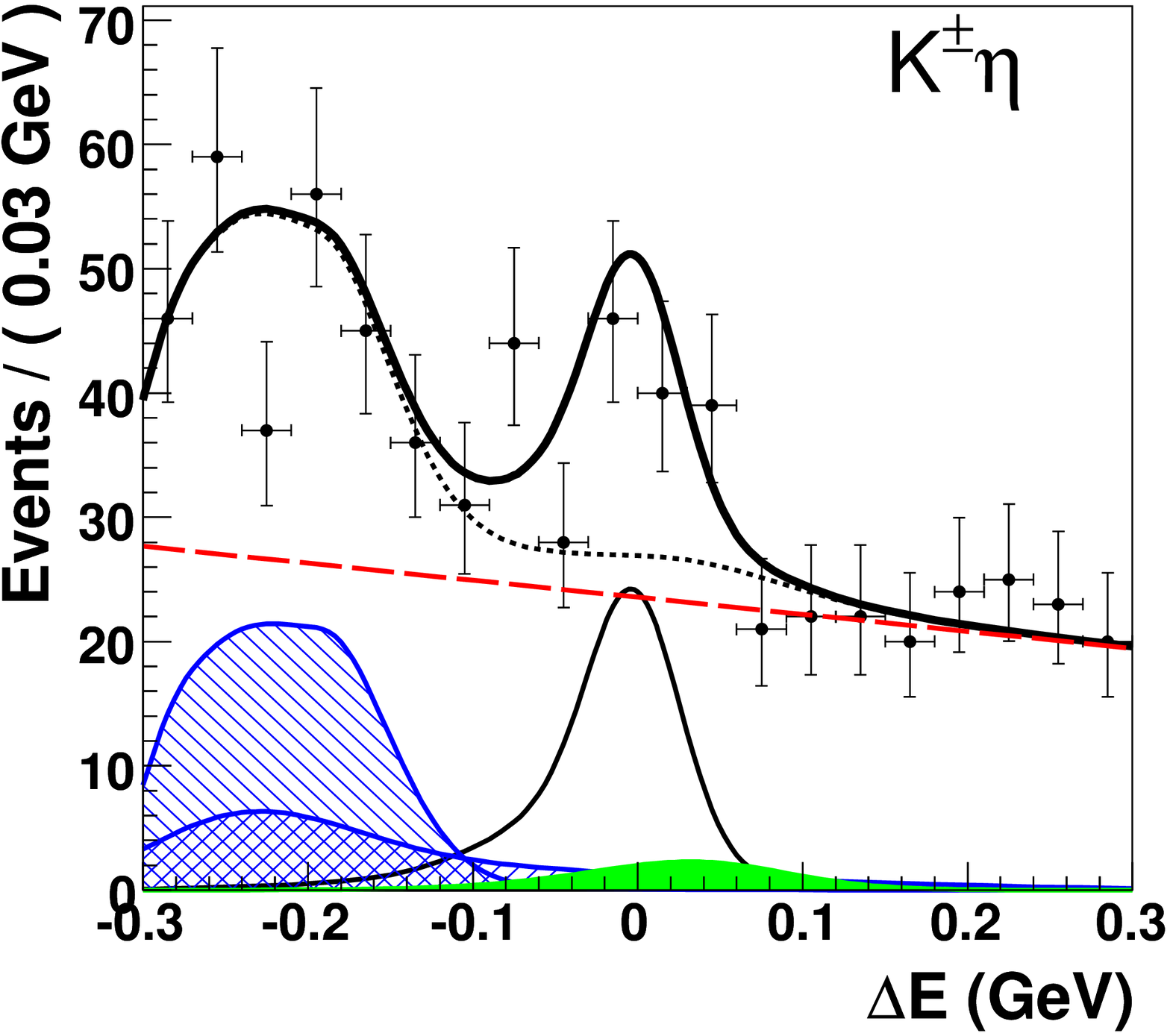} \\ 
   \includegraphics[width=0.30\textwidth]{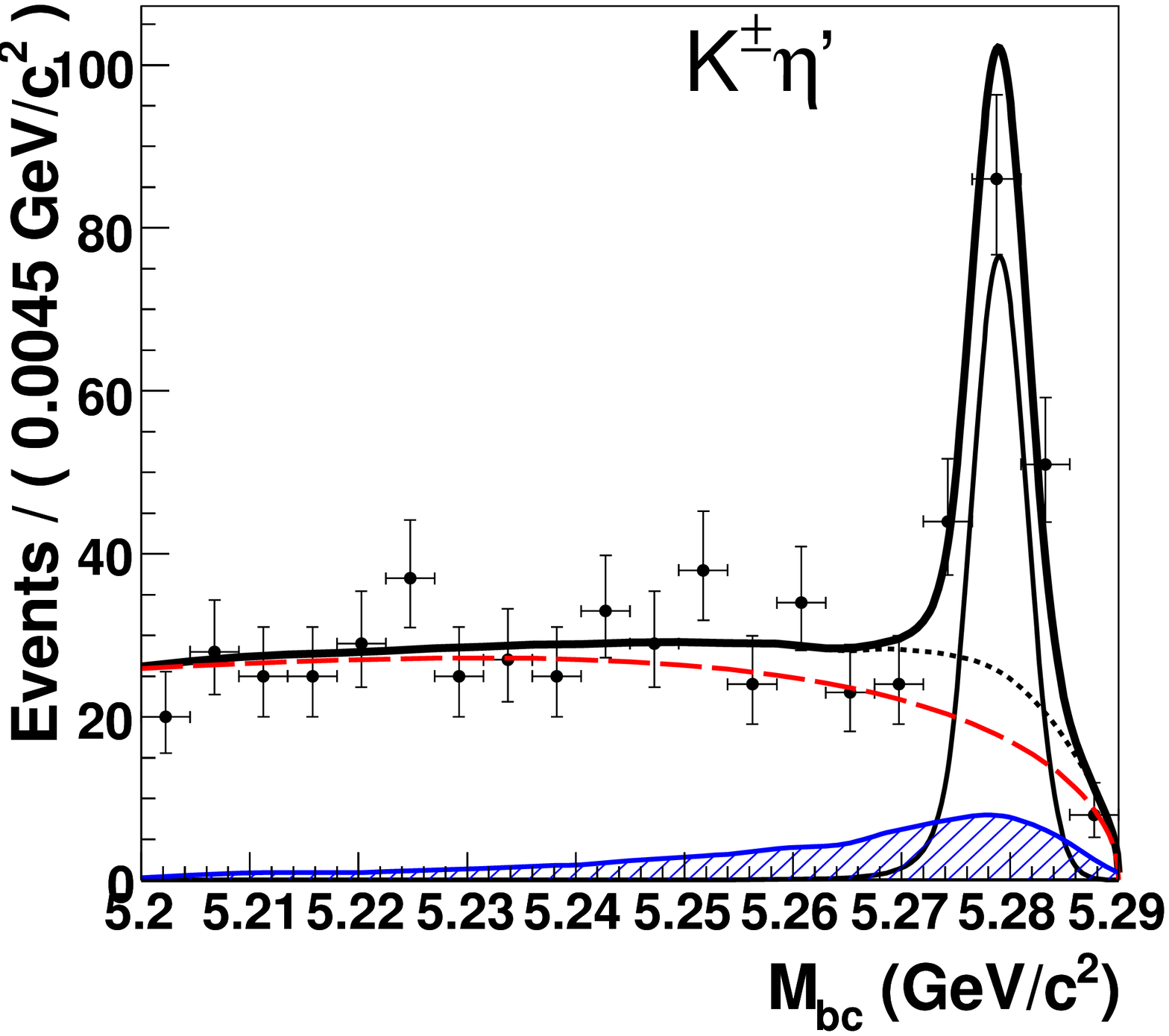} & 
   \includegraphics[width=0.30\textwidth]{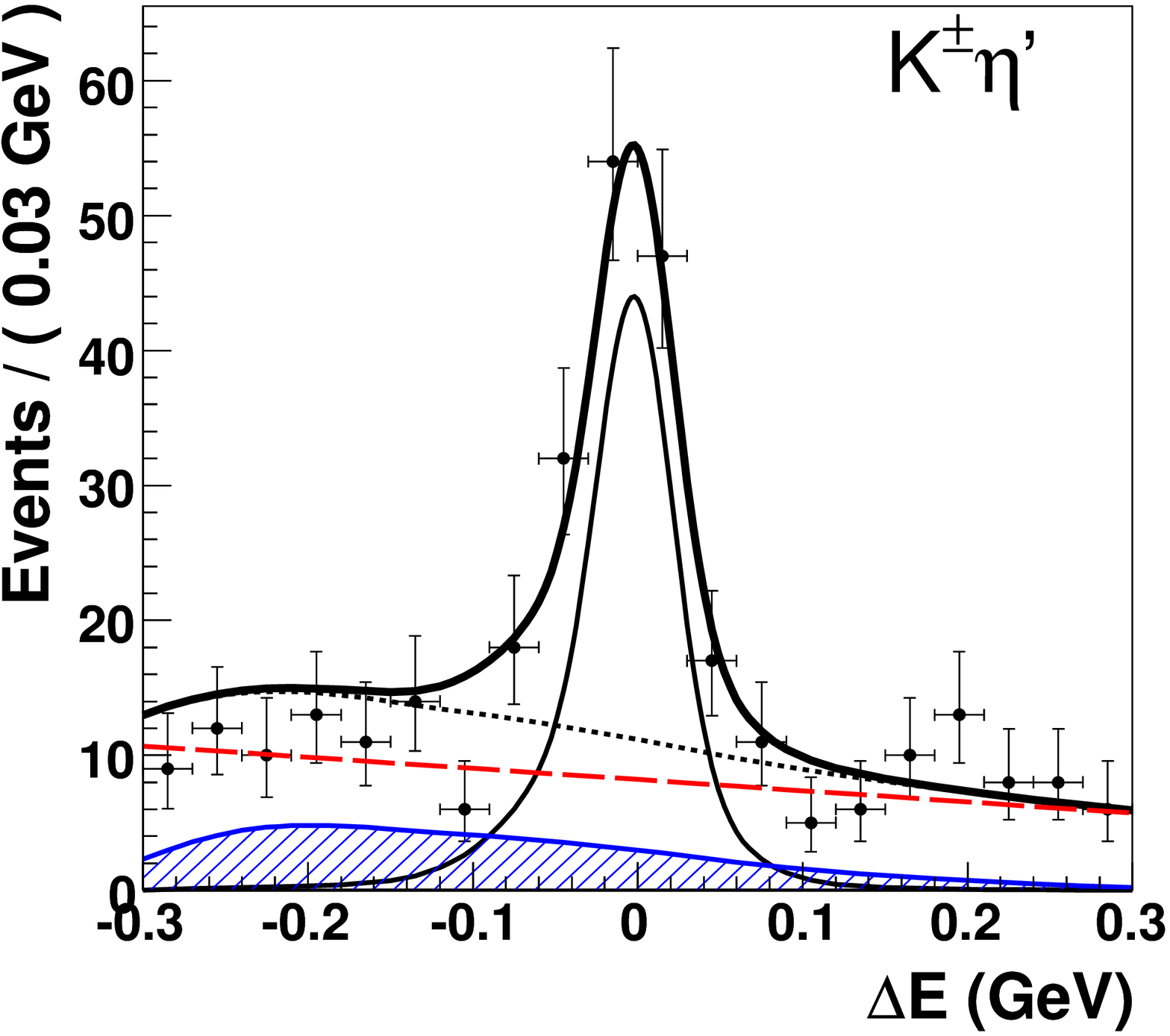} \\ 
   \includegraphics[width=0.30\textwidth]{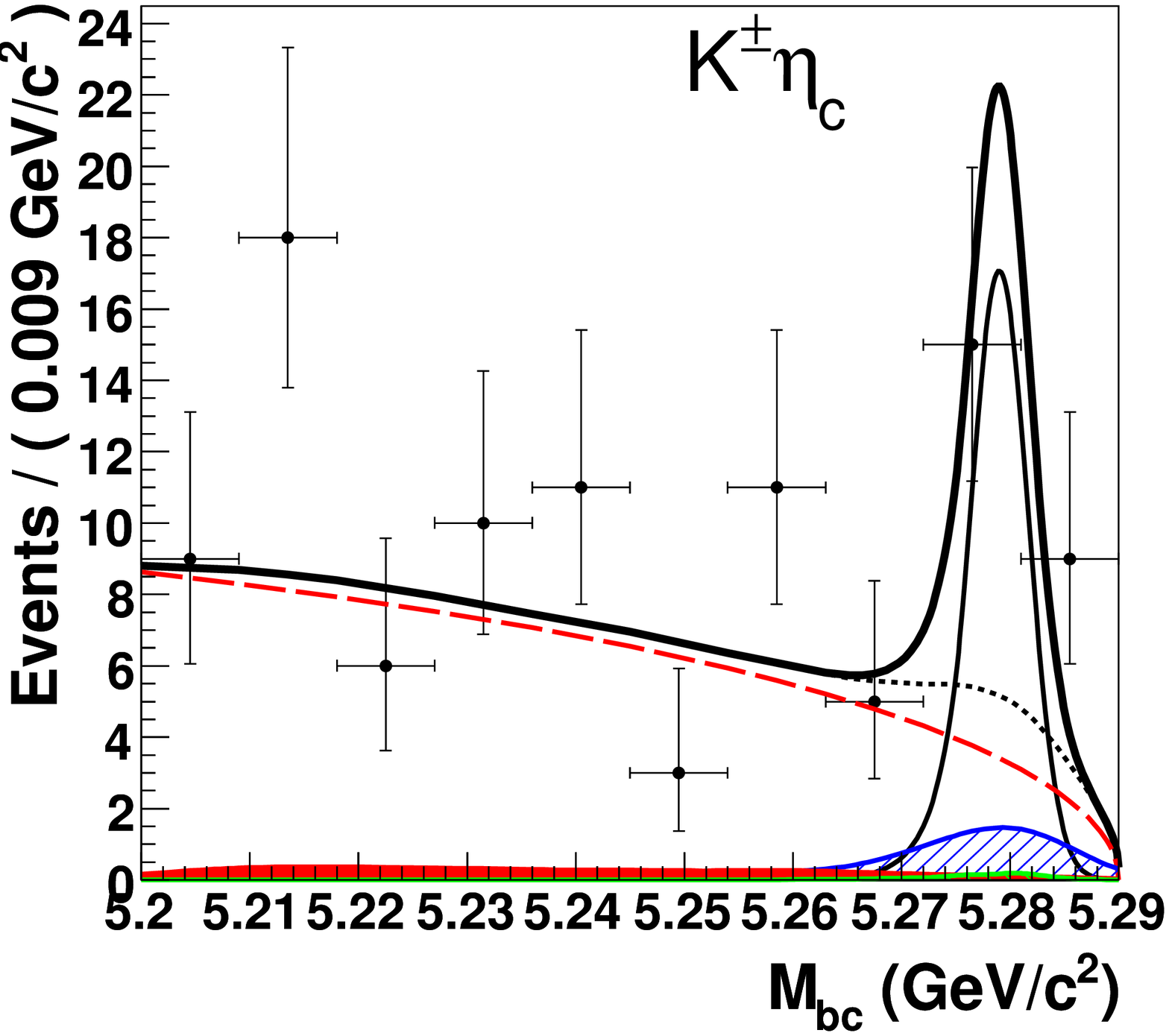} & 
   \includegraphics[width=0.30\textwidth]{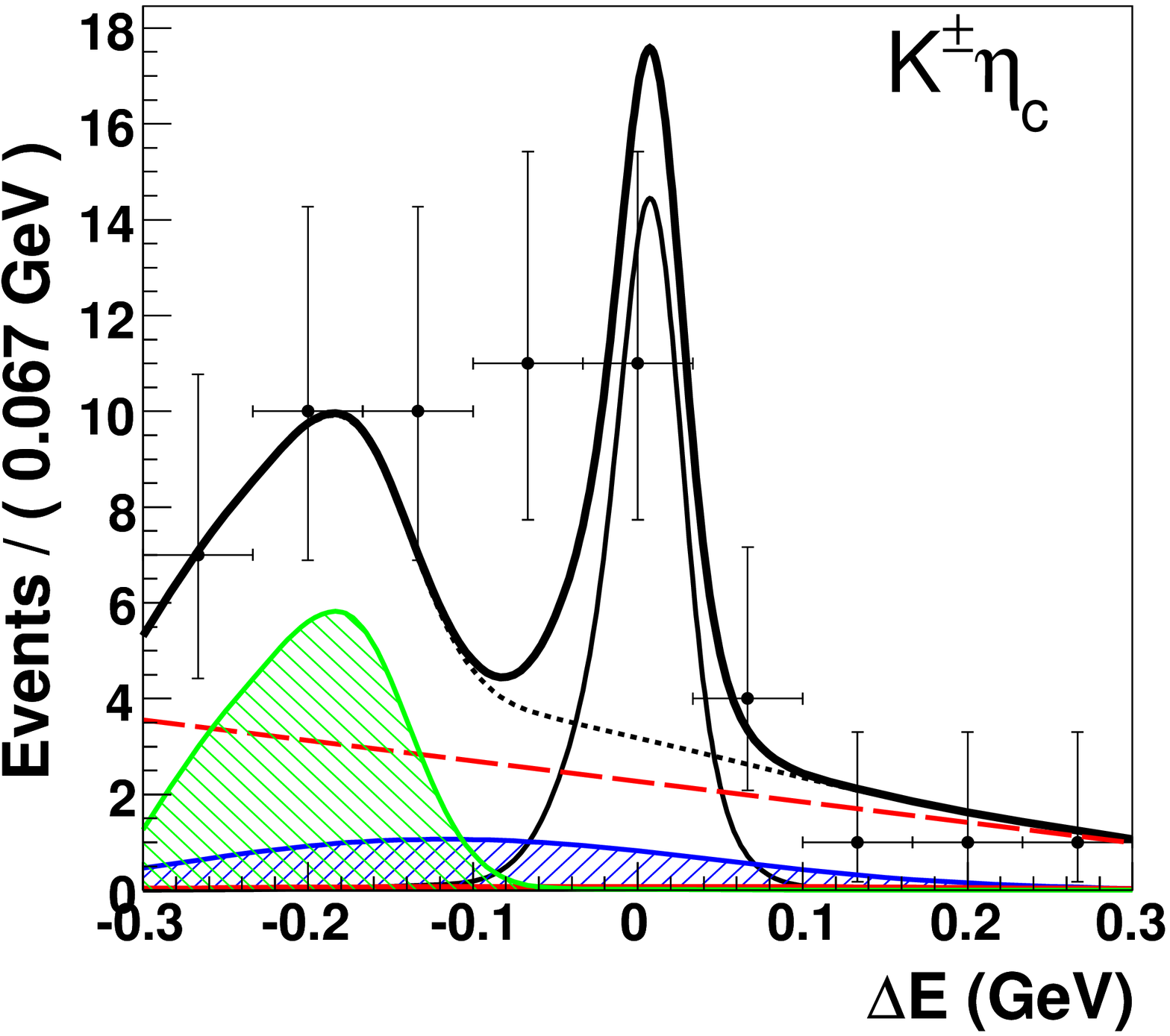} \\
   \includegraphics[width=0.30\textwidth]{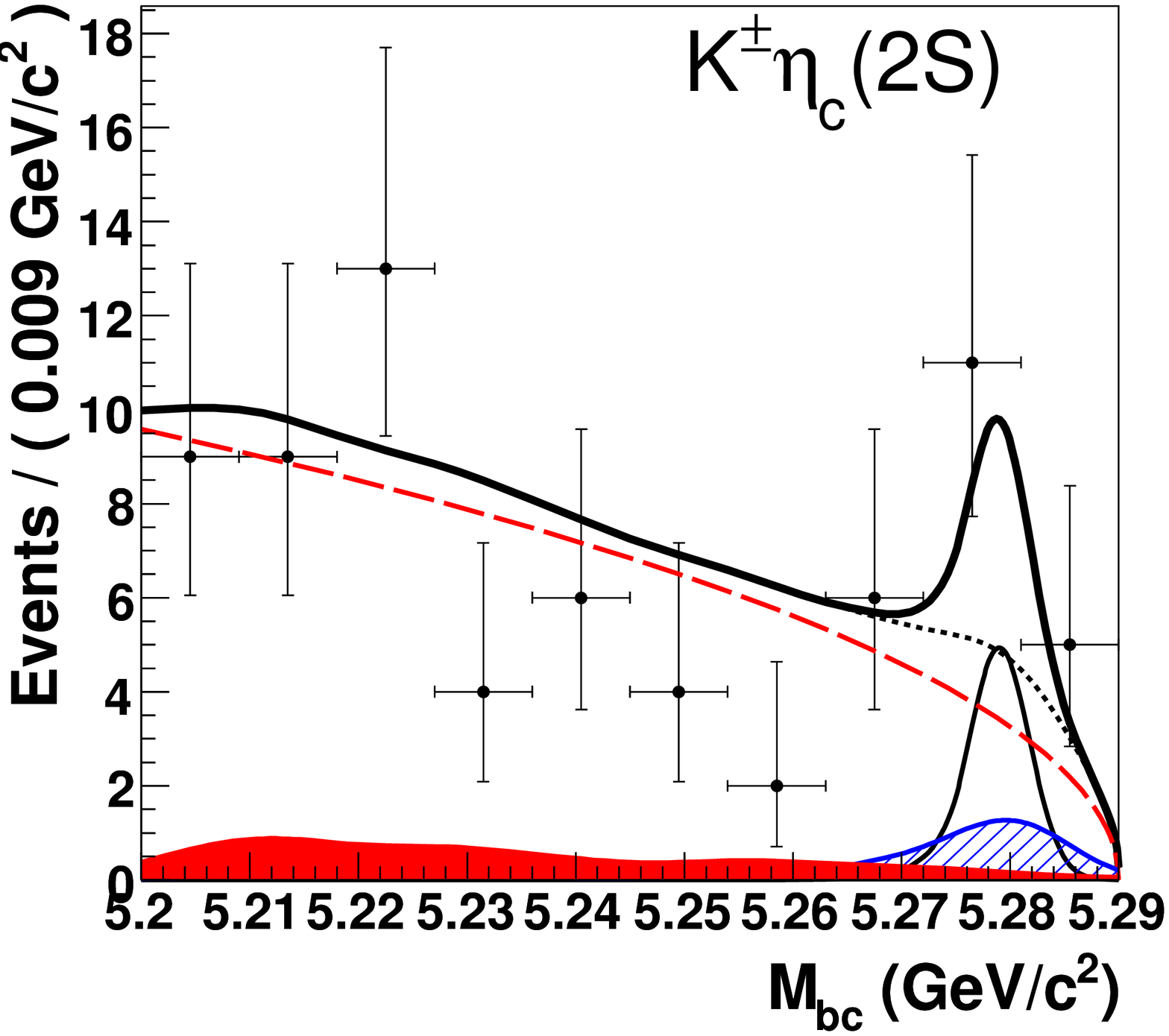} & 
   \includegraphics[width=0.30\textwidth]{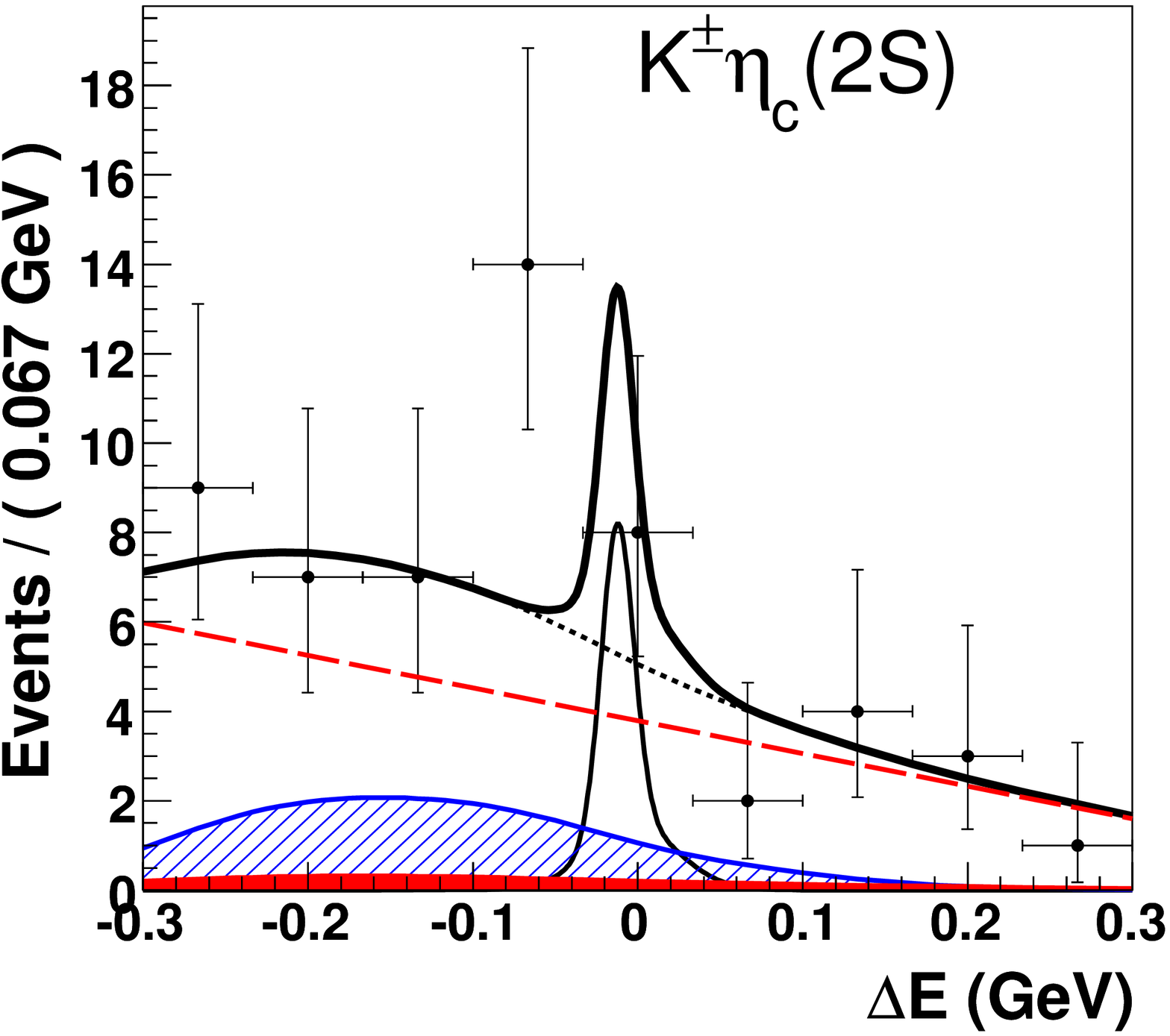} \\
 \end{tabular}
\caption{\mbc\ and \deltae\ projections together with fit results. The first row presents the \BtoKeta\ mode, the second one \BtoKetap , the third one \BtoKetac\ and the last one \BtoKetacp. The points with error bars represent data, the thick solid curves are the fit functions, the thin solid curve is the signal function, the dashed curves show the continuum contribution and the dotted curves show the sum of all background contributions. 
The hatched area present in the whole \deltae\ region is the contribution from the charmless $B$ decays. The hatched area around $\deltae = -0.2\;\gev$ in \BtoKeta\ (\BtoKetac) shows the contribution from \BtoKsteta\ decays (\BtoKstetac). The filled area around $\deltae = 0.05\;\gev$ in the \BtoKeta\ plot is the contribution from $B^{\pm} \to \pi^{\pm} \eta$. The filled area in \BtoKetacp\ is the contribution from the off-time background.}
\label{figure:data_eta}
\end{figure}

\begin{figure}
\centering
 \begin{tabular}{ll}
   \includegraphics[width=0.30\textwidth]{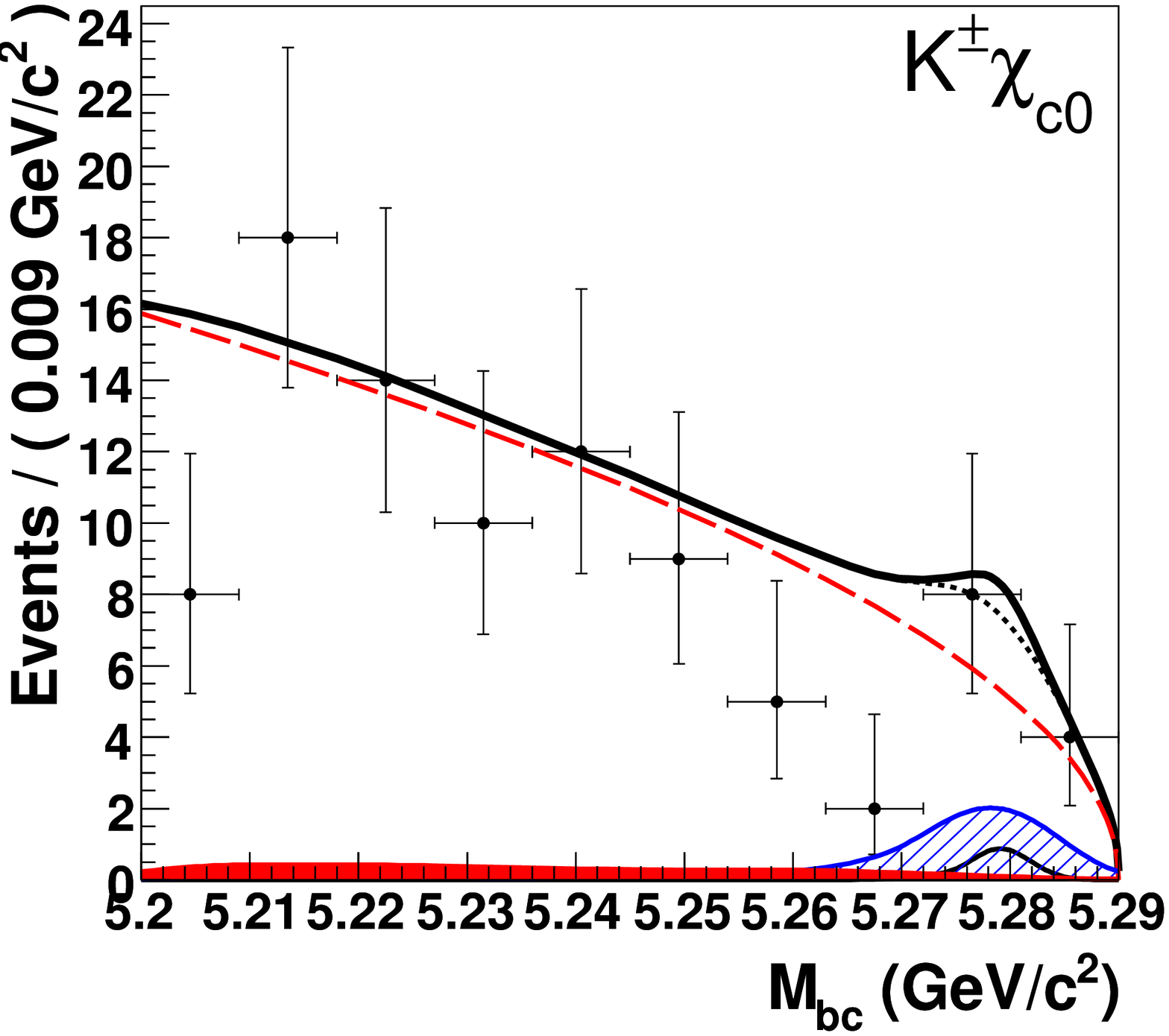} & 
   \includegraphics[width=0.30\textwidth]{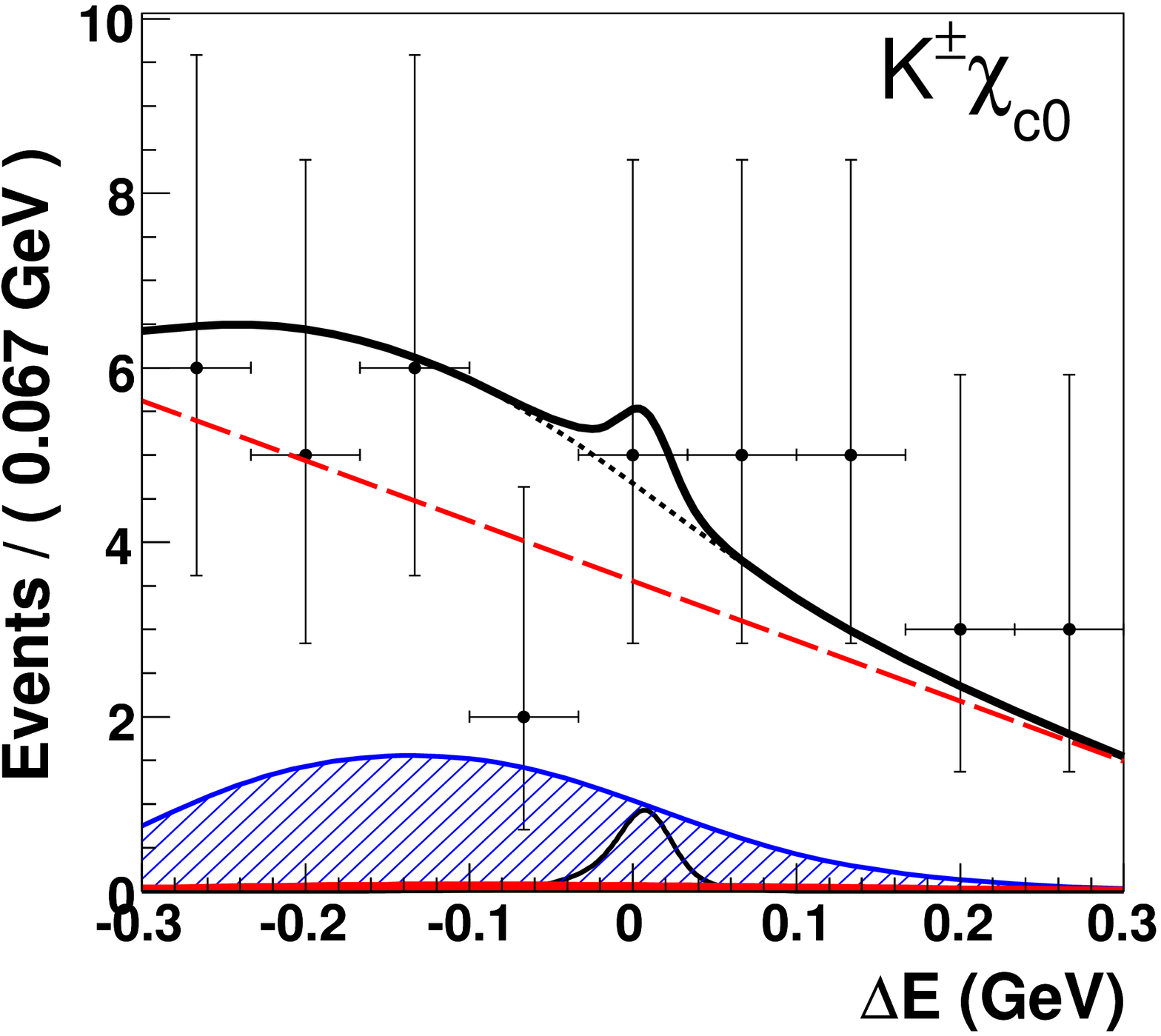} \\ 
   \includegraphics[width=0.30\textwidth]{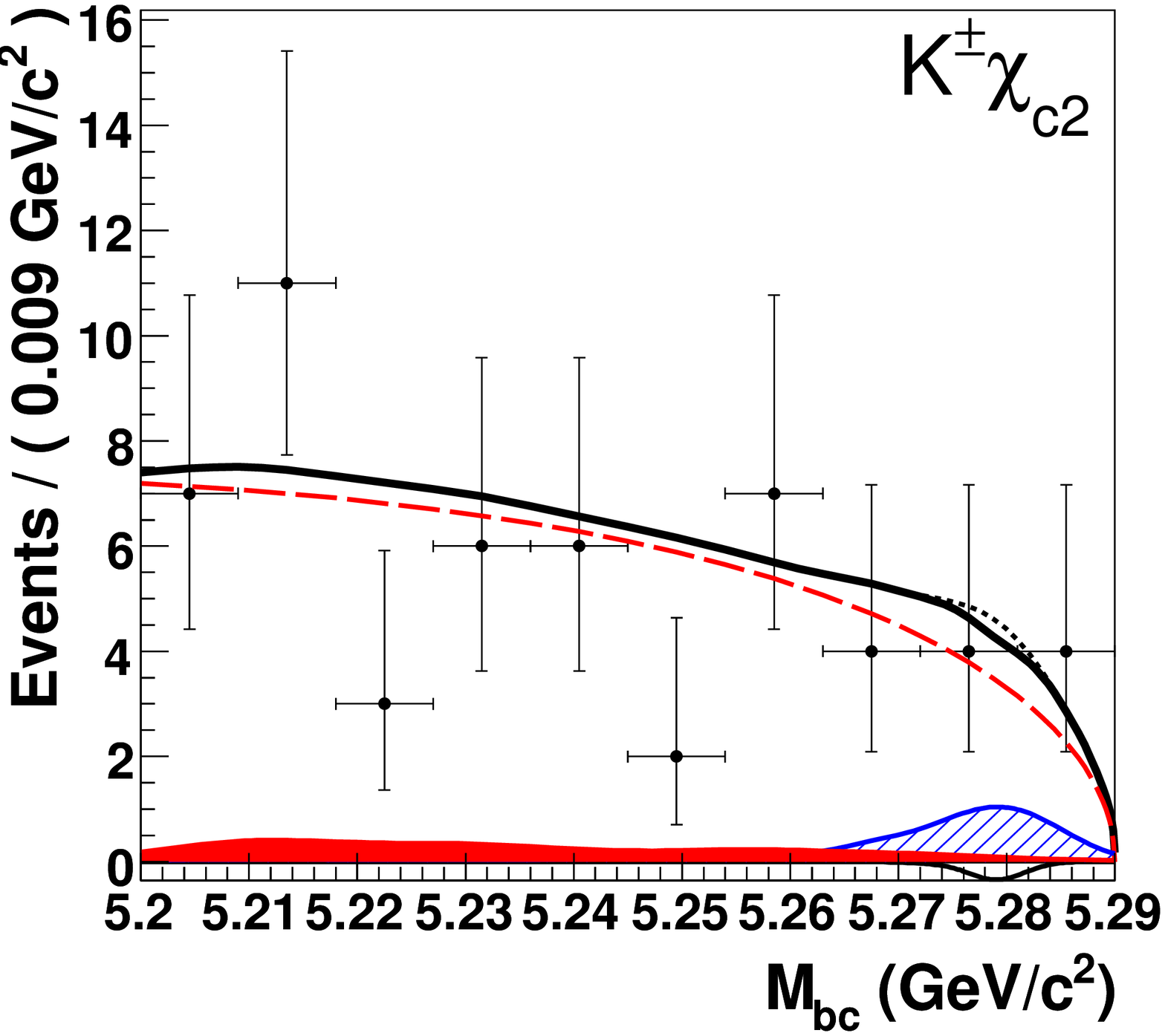} & 
   \includegraphics[width=0.30\textwidth]{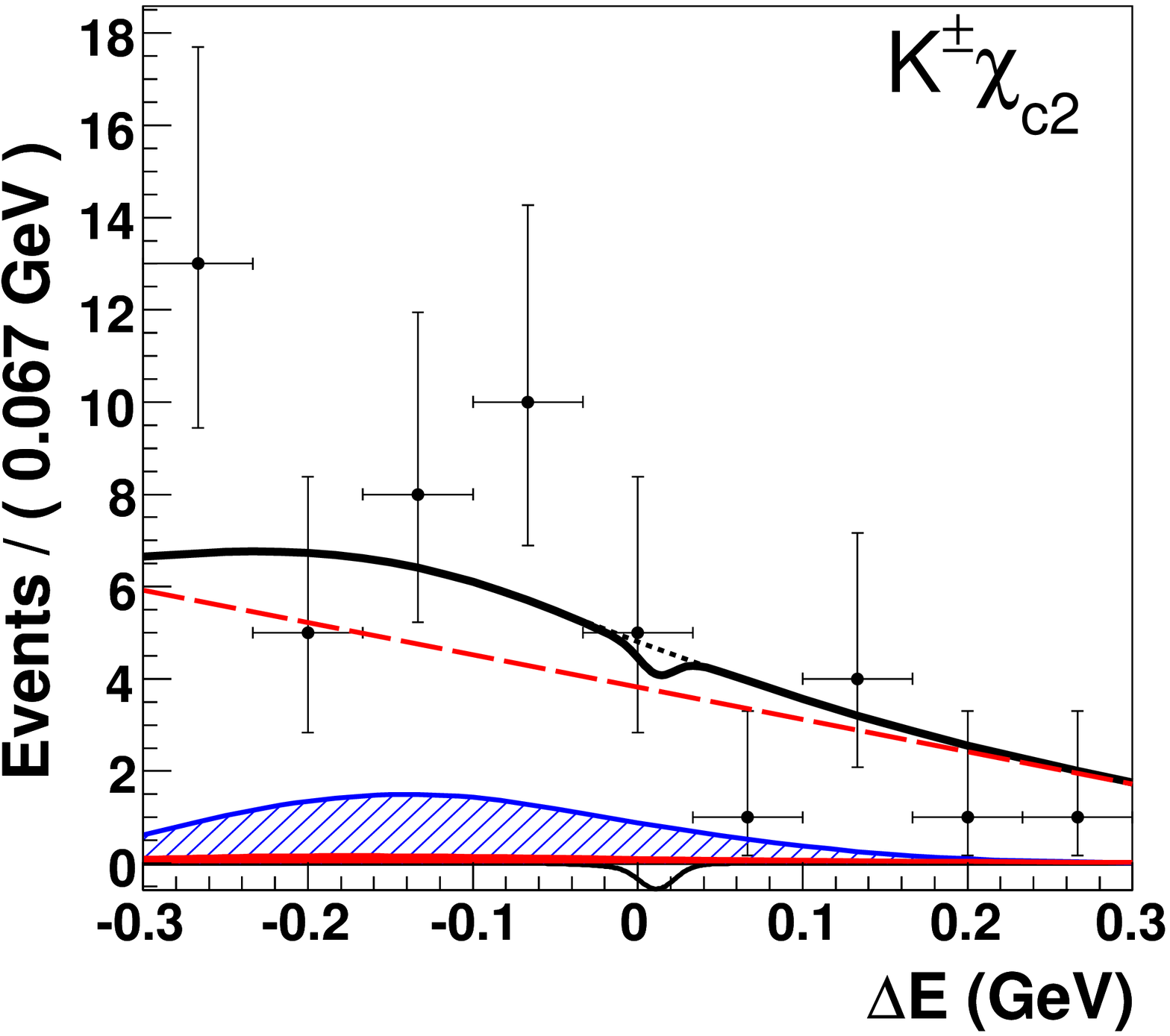} \\ 
   \includegraphics[width=0.30\textwidth]{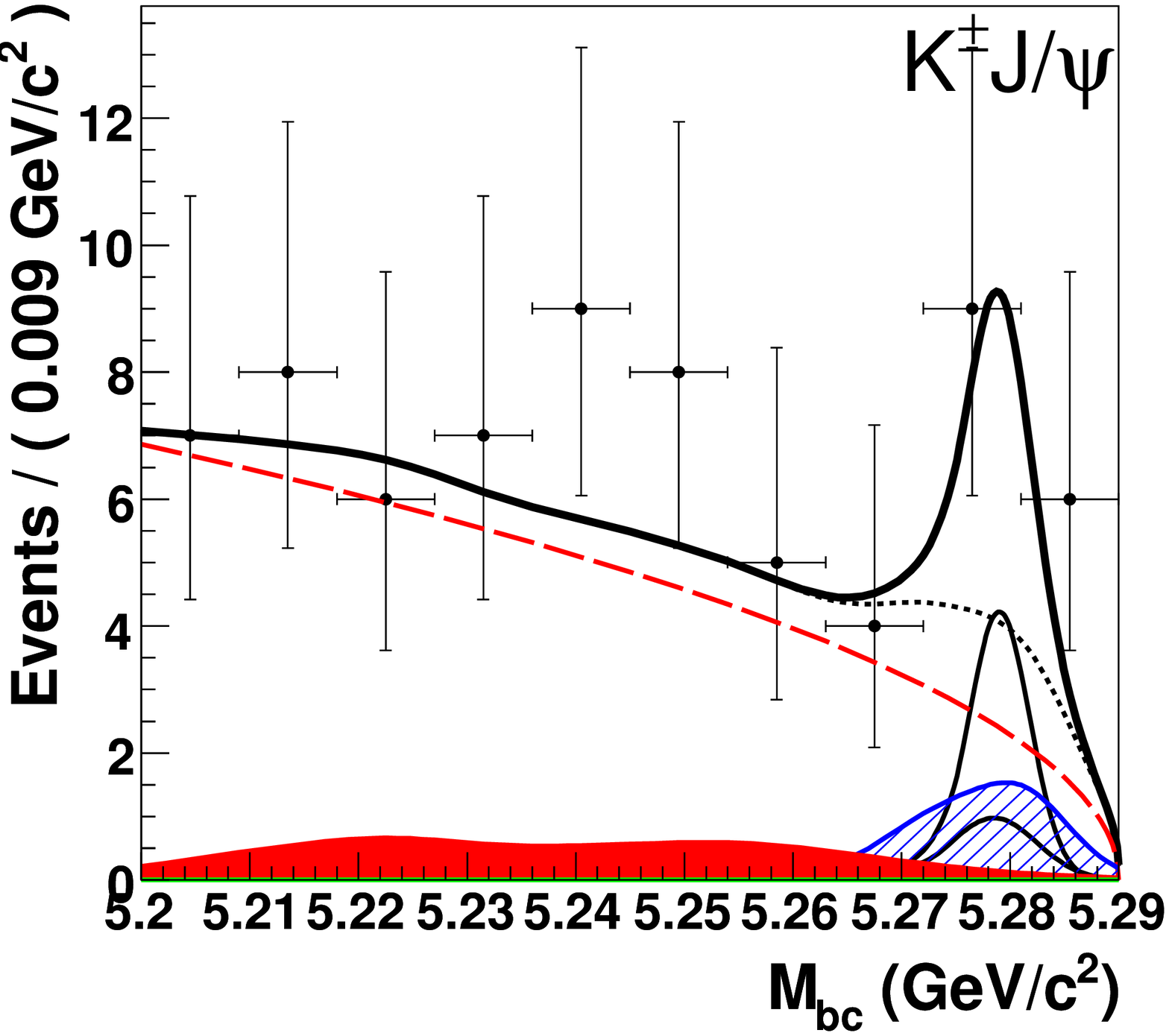} & 
   \includegraphics[width=0.30\textwidth]{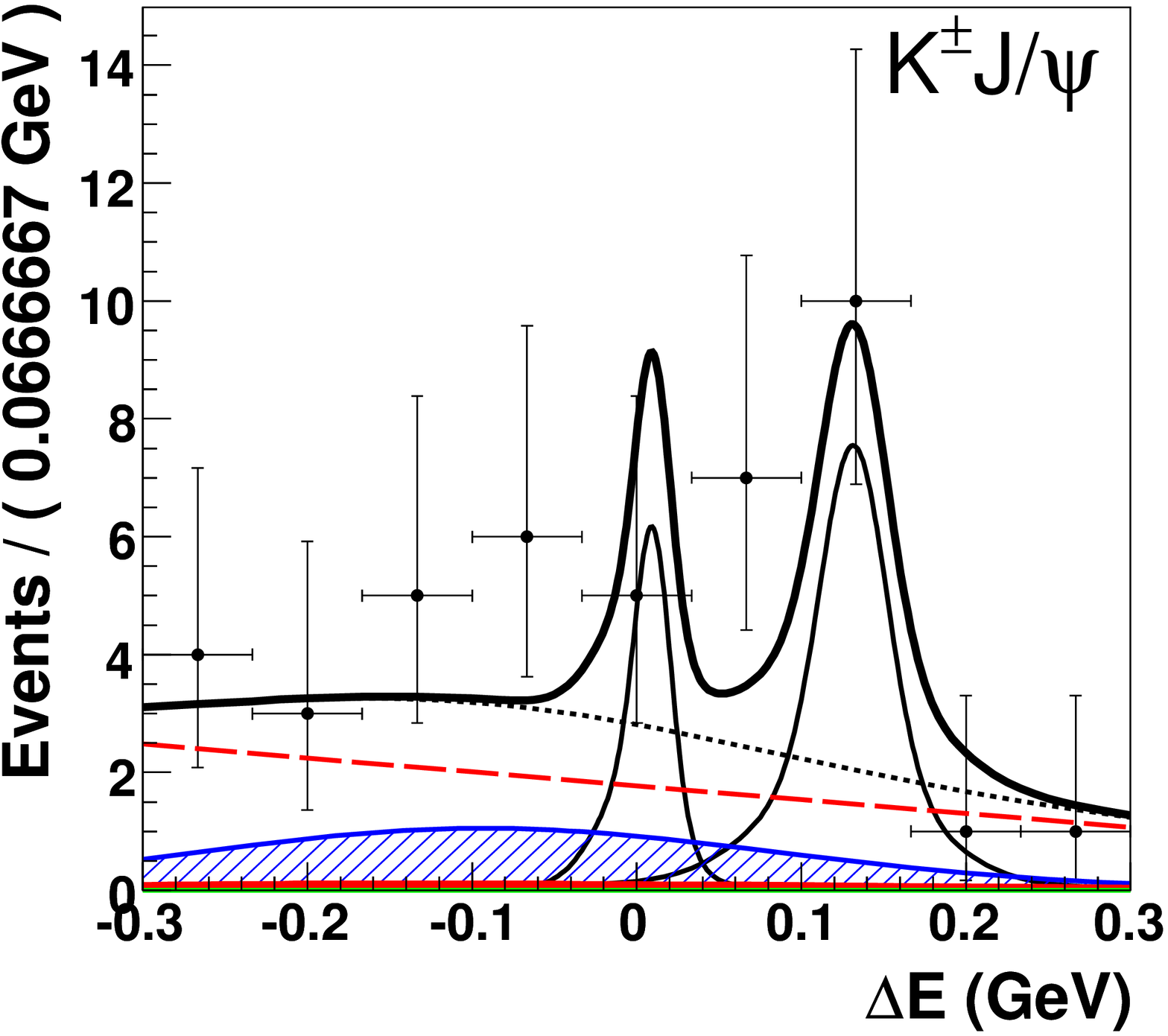} \\
   \includegraphics[width=0.30\textwidth]{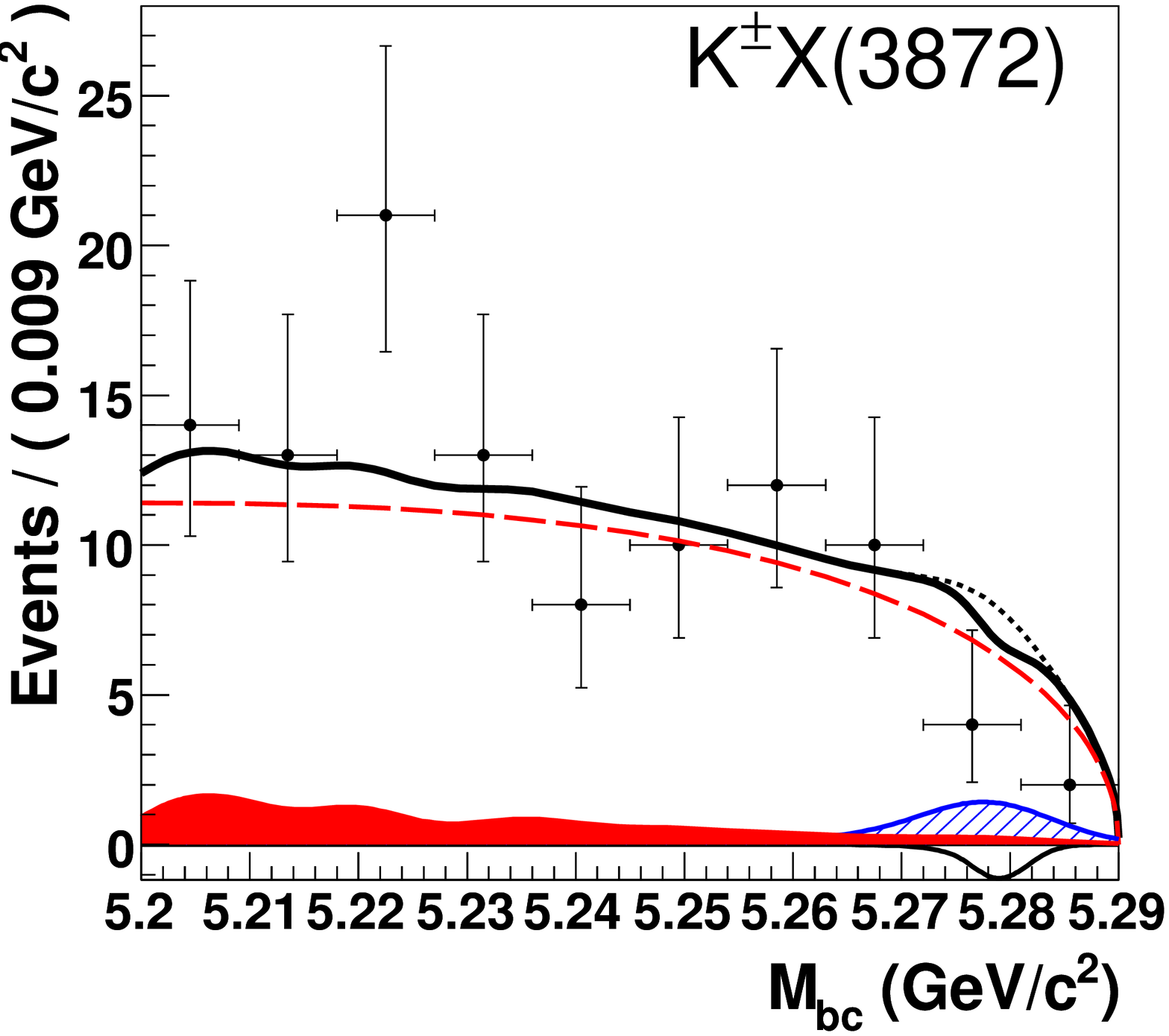} & 
   \includegraphics[width=0.30\textwidth]{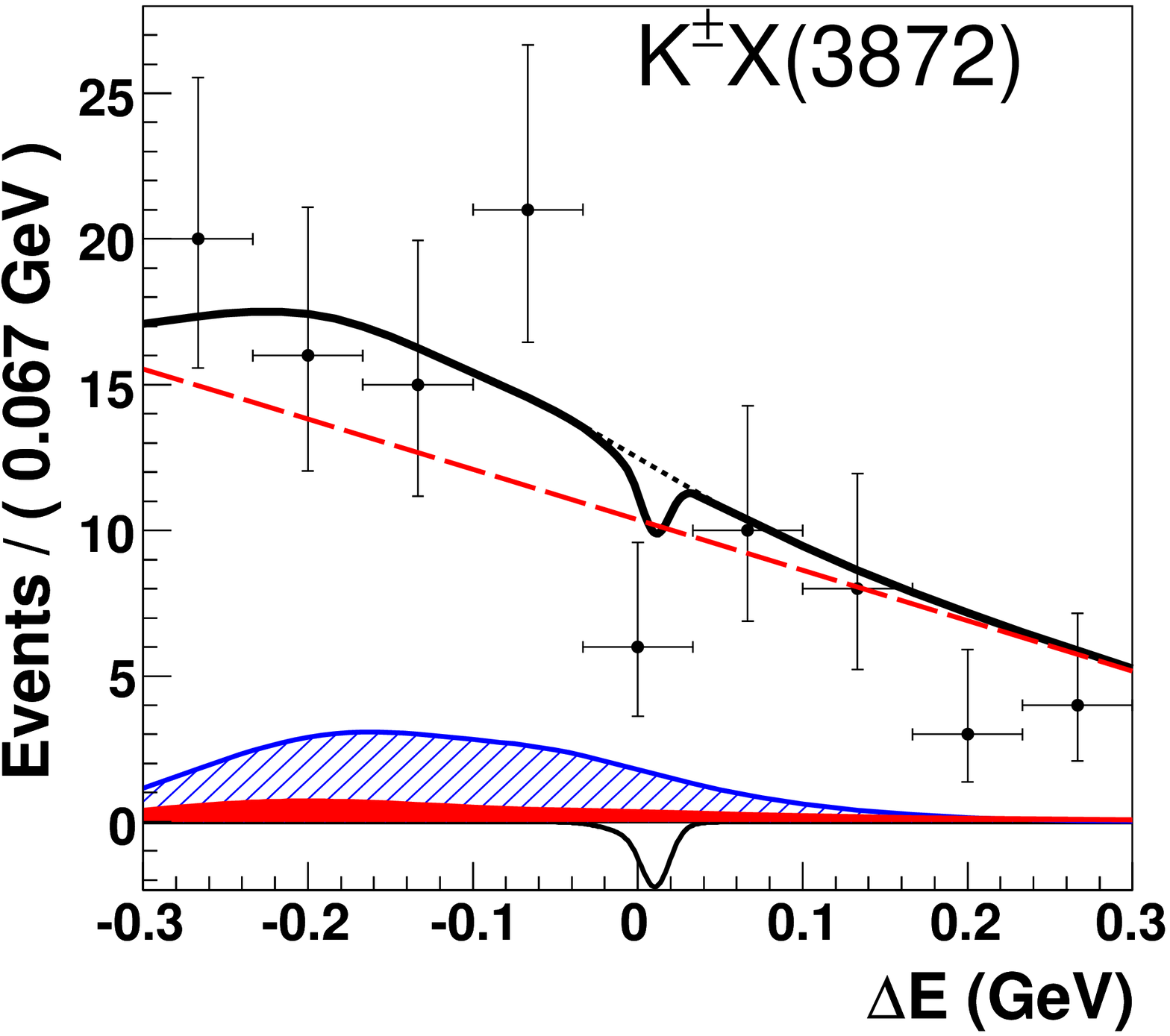} \\
 \end{tabular}
\caption{\mbc\ and \deltae\ projections together with fit results. The first row presents the \BtoKchicz\ mode, the second one \BtoKchict, the third one \BtoKjpsi\ and the last one \BtoKXt. The points with error bars represent data, the thick solid curves are the fit functions, the thin solid curve is the signal function, the dotted curves show the sum of all background contributions, the dashed curves show the continuum contribution, the hatched areas are the contribution from the charmless $B$ decays and the filled areas the contribution from the off-time background. In the \BtoKjpsi\ plots, the \BtoKetac\ cross-feed is visible in the thin solid curves as a small peaking background in \mbc\ that is concentrated around 120 \mev\ in \deltae.
}
\label{figure:data_chicz}
\end{figure}


\begin{thebibliography}{99}

\bibitem{x3872-belle}
S.-K.~Choi, S.L.~Olsen {\it et al.} (Belle Collab.), Phys. Rev. Lett. {\bf 91}, 262001 (2003).

\bibitem{x3872-cdf}
D.~Acosta {\it et al.} (CDF Collab.), Phys. Rev. Lett. {\bf 93}, 072001 (2004).

\bibitem{x3872-d0}
V.M.~Abazov {\it et al.} (D0 Collab.), Phys. Rev. Lett. {\bf 93}, 162002 (2004).

\bibitem{x3872-babar}
B.~Aubert {\it et al.} (BaBar Collab.), Phys. Rev. D {\bf 71}, 071103 (2005).

\bibitem{x3872-1ppor2pp-belle}
K.~Abe {\it et al.} (Belle Collab.), arXiv:hep-ex/0505038 (2005).

\bibitem{x3872-1pp-belle}
G.~Gokhroo, G.~Majumder {\it et al.} (Belle Collab.), Phys. Rev. Lett. {\bf 97}, 162002 (2006).

\bibitem{x3872-1pp2mp-cdf}
A.~Abulencia {\it et al.} (CDF Collab.), Phys. Rev. Lett. {\bf 98}, 132002 (2007). 

\bibitem{x3872-dstd-babar}
B.~Aubert {\it et al.} (BaBar Collab.),	arXiv:0708.1565 (2007).

\bibitem{x3872-ceq1-cdf}
A.~Abulencia {\it et al.} (CDF Collab.),  Phys. Rev. Lett. {\bf 96}, 102002 (2006).

\bibitem{x3872-ceq1-belle}
K.~Abe {\it et al.} (Belle Collab.), arXiv:hep-ex/0505037 (2005).

\bibitem{x3872-ceq1-babar}
B.~Aubert {\it et al.} (BaBar Collab.), Phys. Rev. D {\bf 74}, 071101 (2006).

\bibitem{jpsiforbidden}
L.D.~Landau, Dokl. Akad. Nauk USSR {\bf 60}, 207 (1948) and Phys. Abstracts A52, 125 (1949).

C.N.~Yang, Phys.\ Rev.\ {\bf 77}, 242 (1950).

\bibitem{jpsiul}
R.~Brandelik {\it et al.} (DASP Collab.), Z.\ Phys.\  C {\bf 1}, 233 (1979).

\bibitem{PDG2006}
W.-M.~Yao {\it et al.} (Particle Data Group), J.\ Phys.\ G {\bf 33}, 1 (2006) and 2007 partial update for the 2008 edition.

\bibitem{schietinger}
M.~Knecht and T.~Schietinger,
Phys.\ Lett.\ B {\bf 634}, 403 (2006).

\bibitem{superbelle}
K.~Abe {\it et al.}, KEK Report 04-4 (2004).

A.G.~Akeroyd {\it et al}, arXiv:hep-ex/0406071 (2004).

\bibitem{superb}
M.~Bona {\it et al.}, arXiv:0709.0451 (2007).

\bibitem{b2kgg}
G.~Hiller and A.S.~Safir, JHEP {\bf 0502}, 011 (2005).

See also: S.R.~Choudhury, G.C.~Joshi, N.~Mahajan and B.H.J.~McKellar, Phys. Rev. D {\bf 67}, 074016 (2003); erratum: Phys. Rev. D {\bf 72}, 119906 (2005). 

\bibitem{KEKB}
S.~Kurokawa and E.~Kikutani, Nucl.\ Instr.\ and Meth.\ A {\bf 499}, 1 (2003)
and other papers included in this volume.

\bibitem{Belle}
A.~Abashian {\it et al.} (Belle Collab.),
Nucl.\ Instr.\ and Meth.\ A {\bf 479}, 117 (2002).

\bibitem{svd2} Z.~Natkaniec {\it et al.} (Belle SVD2 Group), Nucl. Instr. and Meth. A {\bf 560}, 1 (2006).

\bibitem{SFW}
 The Fox-Wolfram moments were introduced in
 G.C.~Fox and S.~Wolfram, Phys.\ Rev.\ Lett.\ {\bf 41}, 1581 (1978).
 The Fisher discriminant used by Belle, based on modified Fox-Wolfram
 moments, is described in 
 K.~Abe {\it et al.} (Belle Collab.), Phys.\ Rev.\ Lett.\ {\bf 87},
 101801 (2001) and
 K.~Abe {\it et al.} (Belle Collab.), Phys.\ Lett.\ B {\bf 511}, 151
 (2001). 

\bibitem{TaggingNIM}
H.~Kakuno {\it et al.}, Nucl.\ Instr.\ and Meth.\ A {\bf 533}, 516 (2004). 

\bibitem{crystalball}
J.E.~Gaiser {\it et al.} (Crystal Ball Collab.),
Phys.\ Rev.\ D {\bf 34}, 711 (1986).

\bibitem{argus}
H.~Albrecht {\it et al.} (ARGUS Collab.),
Phys.\ Lett.\ B {\bf 185}, 218 (1987).

\bibitem{keyspdf}
K.S.~Cranmer,
Comput.\ Phys.\ Commun.\ {\bf 136}, 198 (2001).

\bibitem{b2keta-belle}
P.~Chang {\it et al.} (Belle Collab.), Phys.\ Rev.\ D {\bf 75}, 071104 (2007).


\end{thebibliography}
\end{document}